\documentclass[12pt,preprint]{aastex}  
\usepackage{graphicx}
\usepackage{epsfig}
\usepackage{float}
\usepackage{natbib}
\usepackage{amsbsy}
\received{2002 August 19}
\begin{document}

\title{Magnetars in the Metagalaxy: An Origin for Ultra High Energy Cosmic Rays in the Nearby
   Universe}
\author{Jonathan Arons\altaffilmark{1}}
\affil{University of California, Berkeley}
\affil{Department of Astronomy, 601 Campbell Hall, Univ. of California, Berkeley, CA. 94720}
\altaffiltext{1}{also Department of Physics and Theoretical Astrophysics Center}
\email{arons@astron.berkeley.edu}

\begin{abstract}

I show that the relativistic winds of newly born magnetars (neutron stars with
petagauss surface magnetic fields) with initial spin 
rates close to the centrifugal breakup
limit, occurring in all normal galaxies with massive star formation, 
can provide a source of ultrarelativistic light ions with an $E^{-1}$ injection spectrum,
steepening to $E^{-2}$ at higher energies, with an upper cutoff at
$10^{21}-10^{22}$ eV. Interactions with the CMB yield a spectrum at
the Earth which compares favorably with the spectrum of Ultra-High Energy Cosmic
Rays (UHECR) observed at energies up to a few$\times 10^{20}$ eV. The fit to the 
observations suggests that $\sim 5-10$\% of the magnetars are
born with rotation rates and voltages sufficiently high to allow the acceleration of the UHECR. 
The form the spectrum incident on the Earth takes depends sensitively on the mechanism and 
the magnitude of gravitational wave losses 
during the early spindown of these neutron stars - pure electromagnetic spindown (the $E^{-1}$
injection spectrum) yields a GZK feature (a flattening of the $E^3 J(E)$ spectrum) below
$10^{20}$ eV, rather than a cutoff, while a moderate GZK cutoff appears if gravitational 
wave losses are strong enough to steepen the injection spectrum above $10^{20}$ eV. The flux above 
$10^{20}$ eV comes from magnetars in relatively nearby galaxies ($D < 50$ Mpc.)  

I outline the probable physics of acceleration of such particles in a magnetar's wind - it is a
form of ``surf-riding'' in the approximately force free fields of the wind. I also
show how the high energy particles can escape with small energy
losses from the magnetars' natal supernovae. In particular, I show that the electromagnetic energy
emitted by the magnetar ``shreds'' the supernova envelope in times short enough to allow 
most of the relativistic energy to escape largely unmimpeded into the surrounding
interstellar medium, where it drives a relativistic blast wave that expands to parsec scale
before slowing down to nonrelativistic speeds. I also show that since the ions are accelerated
in a region where the magnetic field has the structure of a strong electromagnetic wave but 
propagate at larger radii through a region of weaker magnetic field near the rotational equator
of the outflow, the ultrahigh energy
particles escape with negligible adiabatic and radiation losses. 

The requirement that
the magnetars' relativistic winds not overproduce interstellar supershells and unusually
large supernova remnants suggests that
most of the initial spindown energy is radiated in khz gravitational waves for several 
hours after each supernova. For typical distances to events which contribute to 
$E > 100 $ EeV air showers, the model predicts gravitational wave strains 
$\sim 3 \times 10^{-21} $. Such bursts of gravitational radiation should correlate with
bursts of ultra-high energy particles. The Auger experiment should see bursts of particles
with energy above 100 EeV every few years.

\end{abstract}

\section{Introduction}

I study the possibility that relativistic winds from rapidly rotating {\it magnetars},
neutron stars with surface dipole fields on the order of $10^{15}$ Gauss
(Duncan and Thompson 1992, Paczynski 1992, Kouveliotou {\it et al.} 1998, 1999) 
create the
highest energy cosmic rays (ultra-high energy cosmic rays, a.k.a. UHECR).  
I assume magnetars occur in all normal galaxies which form massive stars, with
the UHECR ariving from outside our own Galaxy, except on the occasions
(perhaps once per $10^5 $ years) when a rapidly rotating magnetar is born in our 
own galaxy. This  model has the virtue of having little difficulty in
accelerating protons to energies in excess of $10^{21}$ eV, with the sources
being in all normal (star forming) galaxies and a luminosity density entirely acceptable
from the point of view of the (very approximately known) rate of formation of
magnetars in our own galaxy, thus offering an explanation of the
puzzling air showers with energies  above $10^{19.6} $ eV without having to introduce
major extrapolations of known or suspected energetics.

In \S \ref{sec:background}, I summarize the data on UHECR,
the loss processes that affect their transport through intergalactic space, 
the energetics of
UHECR and
aspects of the astronomy and physics of magnetized compact objects as known
from studies of rotation powered pulsars that are relevant to the present 
investigation. \S \ref{sec:spectrum} outlines the calculation of the 
particle injection spectrum from a newly born magnetar and the effects
intergalactic transport have in altering the injected spectrum to 
the spectrum received at the Earth. In \S \ref{sec:escape}, I discuss
the escape of the relativistic wind from a magnetar's natal supernova, and in
\S \ref{sec:blast} I outline how the wind drives a relativistic blast wave 
containing a Magnetar Wind Nebula (MWN) into the surrounding interstellar medium. 
The accceleration mechanism of the UHE ions in the magnetar's wind is addressed in
\S \ref{sec:accel}, and their escape from
the expanding MWN nebula is outlined in 
\S \ref{sec:MWNescape}. The effect the electromagnetic energy
lost from a magnetar has in creating HI supershells in the interstellar medium
is discussed in \S \ref{sec:supershells}, with results that are used to suggest
that more than 90\% of a newly born, rapidly rotating 
magnetar's $5 \times 10^{52} $ ergs of rotational
energy gets lost as gravitational radiation. The typical strains of such gravity waves,
from relatively nearby events that could contribute particles in the UHECR spectrum with
energies above $10^{20}$ eV, are estimated in \S \ref{sec:gravity_waves}. The possibility
of observing  bursts of UHECR associated with the birth of individual magnetars
gets attention in \S \ref{sec:beaming}. I discuss the
relation of this study to other work in \S \ref{sec:other_models}.  I draw my conclusions 
in \S \ref{sec:conclusions}.

The most prominent difference between this model for UHECR and most others
which appeal to sources more or less uniformly distributed throughout the Universe
is that the underlying objects create an extremely flat injection spectrum, with 
particles per unit energy range injected at a rate $\propto E^{-1}[1+(E/E_g)^s]^{-1}$, with
$E_g$ reflecting the strength of gravity wave emission on the magnetars' early spin down;
for the model developed in detail here, $s=1$.  As a result, the usual Greisen-Zatsepin-Kuzmin 
(GZK) cutoff disappears. This cutoff, apparent
in models that assume power injection spectra $\propto E^{-s}, s\geq 2$ 
(see Berezinsky {\it et al.} 2002 for a recent example of such ``conservative'' models), becomes a
flattening of the observed $E^3 J(E)$ spectrum between $10^{19}$ and $10^{20}$ eV; the
flatttening  becomes a moderate replica of a GZK cutoff if the gravitational wave 
losses are strong enough.

\section{Background \label{sec:background}}

Recent observations of ultra-high energy cosmic
rays (UHECR), manifested as atmospheric air showers with energy above
$10^{18.8}$ eV, suggest an origin in the metagalaxy (Bird {\it et al.} 1995, 
Yoshida {\it et al.} 1995, Takeda {\it et al.} 1999, Abu-Zayyad {\it et al.} 2002a,b); 
see Nagano and Watson (2000), Watson (2002) 
and Halzen and Hooper (2002) for recent reviews. This 
interpretation relies on three basic facts. 

1) The total cosmic ray energy spectrum hardens above 
$E_a \approx 10^{18.8}$ eV, with isotropic
intensity $J(E) \propto E^{-3.2 \pm 0.05}$ below $E_a$ to 
$J \propto E^{-2.8 \pm 0.2}$ for $E > E_a $, a hardening of the spectrum
known as the cosmic ray ``ankle.'' See Nagano and Watson (2000) for a thorough
discussion of the strong evidence for the spectral hardening at  the highest energies,
including the uncertainties in the value of $E_a$.
Such hardening of the spectrum indicates an origin for
the high energy particles causing the showers at energies above $E_a$
differing from the lower energy particle sources.
Numerically, for $E > E_a $
\begin{equation}
J(E) \approx  10^{-36} \left(\frac{E_a}{E}\right)^{2.8 \pm 0.2} \;
{\rm eV^{-1} cm^{-2} s^{-1} ster^{-1}}.
\label{eq:intensity}
\end{equation}
This flux yields the integral 
intensity above $E_a$,  
$J(>E_a) \approx 3 \times 10^{-18} $ cm$^{-2}$-sec$^{-1}$-ster$^{-1}$, 
corrsponding to particle number density 
$n(>E_a) \approx 1.5 \times 10^{-27}$ cm$^{-3}$ and 
energy density 
$U(>E_a) \approx 2 \times 10^{-8}$ eV cm$^{-3}$.

2) Correlations of event directions on the sky with the 
supergalactic equator may exist at energies above $10^{19.5}$ eV (Uchiori 
{\it et al.} 2000), although the results are also marginally consistent
with isotropy, as are other analyses (Takeda {\it et al.} 1999 and 
references therein.) An extragalactic origin remains consistent with either 
interpretation of the isotropy investigations. The few events with energy above
$10^{20}$ eV say nothing useful about anisotropy.

3) The shower data are consistent with light nuclei or protons as the 
initiating events above $E_a$. At lower energies, in the
cosmic ray ``shin'', lying between the ``knee'' at $10^{15}$ eV and 
the ankle, the data suggest a heavy nuclei composition (Bird {\it et al.} 1994).

If light charged particles such as protons initiate the
UHE showers, the Larmor radii of such particles in the interstellar 
magnetic field,  
$r_L = E/ZeB = 100 (E/10^{20} \;{\mathrm eV}) (10^{-6} \;{ \rm Gauss} / 
Z B )$ 
kpc, exceed the radial extent of the gaseous disk of our galaxy (for $Z = 1$), 
and greatly
exceed the vertical thickness of the disk - indeed, $r_L (100 \; {\rm EeV})$ 
greatly exceeds the thickness of the galactic synchrotron halo.  
Even if $Z = 26$, 
as in the galactic pulsar source model of Blasi {\it et al.} (2000), the 
Larmor radius of 100 EeV particles still exceeds the thickness of the known
magnetized disk (the synchrotron halo, with scale height $\sim 1 $ kpc),
implying a substantial reduction of the flux estimated by Blasi {\it et 
al.}\footnote{From what is known of the top 
layers of a pulsar's atmosphere from X-ray observations, the composition 
almost certainly is not iron. Helium may be a more probable candidate, based on 
interpretaion of neutron stars' X-ray spectra [including line features in
one case, Sanwal {\it et al.} 2002 and references therein, or perhaps O and Ne
(Hailey and Mori 2002).] Lighter elements would have 
floated up to the  X-ray photosphere (Zavlin {\it et al.} 1998, 
Pavlov and Zavlin 2000).
Such high altitude ions (perhaps a few centimeters above the crust) are the reservoir 
for extraction from the star by the 
electric fields parallel to the almost vertical $B$ fields, to be injected into 
the escaping wind, not the 
iron thought to dominate the crustal composition.  Therefore, a charged 
particle model of the UHECR based on acceleration driven by neutron stars
more plausibly should accelerate protons or $\alpha $ particles, as 
indicated by the shower structures at energies above $E_a$, which in turn 
requires consideration of an extragalactic model, to be consistent with the
isotropy data.}  The observed isotropy also argues against the UHECR 
being galactic or local group neutrons, which can survive for a distance 
$\sim 9 (E/10^{18} \; {\rm eV} )$
kpc from their sources; neutrons at 100 EeV certainly cannot arrive 
from the whole local supercluster.

Extragalactic ultra-high energy charged particles cannot arrive from 
truly cosmological distances. As first noted by Greisen (1966) and by
Zatsepin and Kuz'min (1966), interaction of relativistic charged particles 
with the cosmic microwave background (CMB) degrades their energy 
at a rapid rate
through production of $\pi $ mesons when the CMB photons have sufficient 
energy in the rest frame of the cosmic ray. The attenuation length 
for protons due to pion losses (Protheroe and Johnson 1996, Stanev {\it et al.} 2000,
Fodor and Katz 2000, Berezinsky {\it et al.} 2002) can be 
adequately represented as
\begin{eqnarray}
l_{GZK} & = & -c\frac{E}{\dot{E}} = 
    cT_{GZK} \left[1+\left(\frac{E_{G}}{E} \right)^2 \right], \; 
            E_{G}  =  2.8 \times 10^{20} \; {\rm eV},
             \label{eq:gzk_rate} \\
    D_{GZK}  & \equiv & cT_{GZK} =  20 \; {\rm Mpc}, \; 
     T_{GZK} = 6.5 \times 10^7 \; {\rm years}  \label{eq:gzk_time}.
\end{eqnarray}
Expression (\ref{eq:gzk_rate}) gives the largest loss rate
above $ 4 \times 10^{19}$ eV. Below this energy,  photopion losses
are weaker than those due to $e^\pm $ pair
creation and to cosmological
expansion. For the purposes of the present study, pair creation
and cosmic sxpansion can be adequately represented as having constant
attenuation lengths\footnote{More accurately, 
$l_\pm = 500 [1+(E_\pm /E)^{0.4}]$ Mpc for $E \leq E_\pm = 4 \times 10^{19}$
eV, but the slow increase in path length with decreasing energy 
has no effect on the observable aspects
of the model outlined here, at its present level of development.}, 
$l_\pm = 1000$ Mpc, $T_\pm = 3.3$ Gyr, and $l_H = c T_H =4300$ Mpc,
$T_H = 14$ Gyr. 
Therefore, the average loss rate is
\begin{equation}
\frac{\dot{E}}{E} = - \frac{1}{T_{loss}(E)} = -\frac{1}{T_H} - \frac{1}{T_\pm} -
    \frac{1}{T_{GZK}\left[1+\left(\frac{E_{G}}{E} \right)^2 \right]}.
\label{eq:loss-rate}
\end{equation}

Many hypotheses have appeared concerning the origin of 
these air showers. The most puzzling aspect of the problem has been the observation
of air showers at energies so large (in excess of $10^{20}$ eV) 
that charged particles coming directly from
the accelerator must have started no more than 50 Mpc away, due to the radiation
losses in the CMB.  Within this local patch of the Universe, there is a lack
of the most commonly hypothesized accelerators, disks and jets associated with giant black holes
in the most {\it active} galactic nuclei, an idea with a long history going 
back to Lynden-Bell (1969).  

This scarcity has led to a plethora of
explanations. Some are of the ``bottom up'' variety as in this paper, 
based on acceleration of normal charged particles to extreme energies, usually
in objects not otherwise known to exhibit the ability to accelerate extremely high energy particles
({\it e.g.}, Torres {\it et al.} 2002 and references therein). Others are of the ``top-down''
variety, based on the behavior of particles left over from 
the early Universe - see Sarkhar (2002) for a recent review of such 
ideas. The Z-burst model (Singh and Ma 2003 and references therein)
offers a mixed alternative, with extremely high energy charged
particle acceleration 
and ultra high energy neutrino production moved to high redshift
(where hyper-active AGN are more common), while the
high energy protons we observe are created locally 
through interactions between the UHE neutrinos and 
low energy neutrinos left over from the Big Bang (clustered in dark matter halos, 
if the background neutrinos have
sufficient mass). The main point of the model proposed here
is to argue that adequate accelerators may be common in the local Universe, since they may be
in all {\it normal} galaxies with active star formation. In particular, the
accelerators may be associated 
with magnetars, a relatively rare form of core-collapse supernova debris.

The required galactic luminosity in UHECR gives a strong hint
that the UHECR sources have something to do with massive stars and supernovae
in ordinary galaxies. This luminosity is, using 
(\ref{eq:intensity}) and (\ref{eq:loss-rate}), 
\begin{equation}
L_g^{UHE} = \frac{4\pi}{c n_g} \int_{E_a}^{3 \times 10^{20} \, {\rm eV}}
 dE E \frac{J(E)}{ T_{loss}(E)} = 0.8 \times 10^{39} n_{g2}^{-1} \: {\rm ergs/sec}.
  \label{eq:UHE_gal_lum}
\end{equation}
Here $n_g = 0.02\: n_{g2}$/Mpc$^3$ is the space density of
galaxies (Blanton {\it et al.} 2001), the majority of which actively form stars 
and are not in large clusters.
The particles' intergalactic residence time, weighted by the energy in the observed spectrum in 
(\ref{eq:UHE_gal_lum}),
is $\langle T_{loss} \rangle \approx T_H /6.9 \approx 2.1$ Gyr; when weighted
by particle number, the mean residence time is $\sim T_H /3.8 = 3.7$ Gyr. This luminosity per
galaxy coresponds to the metagalactic luminosity density $5 \times 10^{44}$ ergs/Mpc$^3$-year.
These energetics are consistent with a
supernova related source, as is the case with normal cosmic rays.  The high energy per particle
forms the distinctive peculiarity of the UHECR. 

The basic model for the accelerator proposed 
here is a variant of that advanced
by Blasi {\it et al.} (2000), who assumed the relativistic winds from
pulsars in our own galaxy could be the 
acceleration site, granted that these stars begin their lives as almost
maximally rapid rotators. Relativistic magnetic rotators have magnetospheric
voltage drops across the magnetic field with magnitude 
\begin{equation}
\Phi_{\rm mag} = \frac{\Omega^2 \mu }{c^2} = 
  \Psi_{\rm mag} \frac{\Omega}{c} 
   = 3 \times 10^{22} \frac{\mu}{10^{33} \; {\rm cgs}}
        \left(\frac{\Omega}{10^4 \; {\rm s}^{-1}} \right)^2 \; {\rm Volts},
\label{eq:voltage}
\end{equation}
which is in excess of the energy/charge exhibited in the UHECR,
if the underlying star has large angular velocity ($\Omega > 10^3$ s$^{-1}$) and large
dipole moment ($\mu > 10^{32} $ cgs, corresponding to a surface
dipole field above $10^{14}$ Gauss)\footnote{Indeed, a  
neutron star which had magnetic energy in the poloidal field comparable to the 
gravitational binding energy of the star ($B_{\rm dipole} \sim 
10^{18} $ Gauss) would be able to accelerate charged particles to
$\sim 10^{25}$ eV, if it rotated close to breakup ($\Omega > 10^4 \; {\rm sec}^{-1} $).
Of course, such an object would look nothing like a sphere and would lose
most of its rotational energy in gravitational waves (see below), 
should it be so fortunate to survive the host of instabilities plaguing its existence.
Magnetized fallback disks ({\it e.g.}, Macfadyen and Woosley 1999) 
might be functional equivalents
of magnetically supported, rapidly rotating neutron stars; see \S \ref{sec:other_models}.}. 
Here $\Psi_{\rm mag} = R_L^2 B(R_L)$
is the magnetic flux in the ``open'' field that extends beyond
the light cylinder, located at $r = R_L \equiv c/\Omega = 30 / \Omega_4$ km
that connects the 
magnetosphere to the outside world, and 
$\Omega =10^4 \Omega_4 \; {\rm s}^{-1} $.

Acceleration in the wind avoids the catastrophic radiation losses
associated with acceleration inside a rapidly rotating neutron star's
magnetosphere.  Charged
particles moving on paths with radius of curvature $\rho_c $ have radiative power
$P = (2/3) (q^2 c /\rho_c^2) (E/mc^2)^4 $. Radiation losses would therefore prevent
protons from accelerating to energies above \newline
$
E_{rad} \approx  10^{16.5} (A^2 /Z)^{1/3} \Omega_4^{-1/3} (r /R_L)^{1/3} 
 (\rho_c /r)^{2/3},
$
eV, where a particle's mass is $Am_p $ and its charge is $Ze $.  
Within and near the magnetosphere, the strong magnetic field guides
particles along $B$. Therefore, even if strong parallel electric
fields can form, the radius of curvature of a particle's orbit
is $\rho_c \sim r \sim R_L $ and all the energy expended in accelerating
the putative cosmic ray gets lost as gamma rays (and $e^\pm$ pairs.) 

At radii $r \gg R_L$, the physical situation differs.
I shall argue that at large radii, acceleration can and does
occur in an angular sector around a magnetar's rotational equator, where
the radii of curvature of highly accelerated ions' orbits do greatly exceed
the local radius $r$ at each point on their orbits (see \S \S \ref{sec:accel} and 
\ref{sec:MWNescape}), so that radiative losses are negligible. 

Relativistic magnetic rotators, appearing
as rotation powered neutron stars and as black holes with magnetized
disks losing rotational energy in jets, generally create {\it non-radiative}
relativistic outflows.  These outflows - relativistic ``winds'' - 
include wound up magnetic fields
and outflowing plasma, which support the ``MHD'' electric field
${\bf E_\perp} = - (1/c) {\bf v} \times {\bf B}$. The 
magnetic field has magnitude $B \approx B(R_L) (R_L /r) $, an estimate
appropriate both to quasi-spherically expanding winds and to 
(current-carrying) jets. At radii
large compared to $R_L$ , the cross field voltage in the wind is ($v/c \approx 1$)
\begin{equation}
\Phi_{wind} \approx rE_\perp(r) = rB(r) = R_L B(R_L) 
       = \frac{\Psi_{\rm mag}}{R_L} = \Phi_{\rm mag} = \frac{\mu}{R_L^2}
       = \frac{\mu \Omega^2}{c^2} 
\label{eq:wind_volts}
\end{equation}
The last two forms of (\ref{eq:wind_volts}) are appropriate for underlying
rotators which have magnetic dipole moments, such as neutron stars, while
expressing $\Phi_{wind} $ in terms of the open magnetic flux $\Psi$
allows application of the same physical considerations  
to outflows from the magnetized disks thought to orbit black holes.

The estimate given by equation (\ref{eq:wind_volts}) assumes the relativistic wind does
not suffer strong radiative dissipation, which is true
in the outflows from rotation powered pulsars, and is also
the case in electromagnetic models of jets from underluminous galactic 
nuclei in radio galaxies, galactic microquasars and gamma ray bursts (GRBs).  
Then the voltage generated by the 
magnetized rotator is available to accelerate particles at radii
$r \gg c/\Omega $ - for protons, particles of energy greater than
$10^{20.3} $ eV could be accelerated at distances $ r > 0.3 (r/\rho_c)^2 $
pc, a dimension easily achievable in the winds from magnetized rotators.
If the acceleration  process approximates that of a linear accelerator,
rather than of a synchrotron, 
as may be the case if ``surf-riding''
acceleration (e.g., Tajima and Dawson 1979, Chen {\it et al.} 2002, \S \ref{sec:accel}) 
applies, then $r/\rho_c \ll 1 $ and radiation losses
are even less of a constraint on the model (see
\S \ref{sec:MWNescape}).

The new aspect of the idea pursued here is to apply the same physics to
magnetars in all galaxies. These objects have no difficulty (in principle) with
accelerating {\it light} nuclei to the required energies, and by hypothesis, being
located in all galaxies, have no
difficulty in supplying particles with an isotropic
distribution of source directions.  The initial spin rates of magnetars
are unconstrained, except from theoretical considerations of the stars' dynamical stability,
which limits the rotational energy to be less than 25\% of their gravitational
energy ($\Omega < 1.2 \times 10^4$ s$^{-1}$ for a uniform sphere). 

Therefore, one can assume that
the stars are born with sub-millisecond initial rotation periods. The limited evidence on the
initial spins of rotation powered galactic pulsars suggests 
some begin their lives
rotating at  less than their maximal rate, {\it if} 
electromagnetic torques dominate their spindown from birth (e.g., Murray {\it et al.}
2002, Kaspi and Helfand 2002). As I will show, other constraints suggest that at
least for magnetars, gravitational wave torques may govern the initial spindown,
which alleviates concerns over whether all such objects might be born as slow rotators - 
in any case, only 5-10\% of the magnetars must be born as rapid rotators, for the UHECR source
model proposed here to be viable. 

Magnetars are of particular interest as an acceleration site 
since a) they exist (Kouveliotou {\it et al.} 1998, 1999; see the summary
in Baring and Harding's 2001 Table 1), and b)
if they rotate fast enough sometime in their lives, 
they easily can have voltage drops well in excess of the energy 
per unit charge observed so far in UHECR (expression \ref{eq:voltage}).  
$\Omega_4 =1$ corresponds to a rotation 
period of 0.63 msec, about half the centrifugal breakup angular velocity
for a $1.4 M_\odot $
neutron star with 10 km radius, while $\mu_{33} =1$ corresponds to a 
polar surface dipole field of $2 \times 10^{15} \mu_{33} R_{10}^{-3}$ 
Gauss, with $R_{10} = R_*/10$ km; these values corrspond to a charateristic voltage
of $3 \times 10^{22}$ V.    
Clearly, rapidly rotating, magnetized neutron stars can
provide the energy/particle seen in UHECR, in principle, so long as the 
wind carrying the voltage survives to distances far from the star.

The energetics of a magnetar acceleration model are simple. Suppose the particles each gain the 
energy 
\begin{equation}
E(\Omega) = q \eta  \Phi_{\rm mag} = q\eta \frac{\Omega^2 \mu}{c^2}
  = 3 \times 10^{21} Z\eta_1 \Omega_4^2 \mu_{33} \; {\rm eV}, \; \eta_1 \equiv \eta /0.1.
\label{eq:particle-energy}
\end{equation} 
Here $ \eta $ is the fraction of the open field line voltage experienced by each
particle on its way from the star to the outside world. 
If the magnetosphere is relativistic,
the usual electromagnetic rate of  rotational energy loss
\begin{equation}
\dot{E}_{EM} = \frac{4}{9} \frac{\Omega_*^4 \mu^2 }{c^3} 
\label{eq:EMloss}
\end{equation} 
applies\footnote{Expression (\ref{eq:EMloss}) is the standard result for
vacuum dipole radiation, after averaging the geometric factor $(2/3) \sin^2 i$ over the
sphere; $i$ is the angle between the 
dipole moment and the rotation axis. All electromagnetic spindown theories are
thought to lead to comparable rates of energy loss, with 
$\dot{E}_{EM} = c\Phi_{\rm mag}^2$.}. 

I adopt the view that electric conduction currents are a major part of the support for
the electromagnetic fields which exert torques on the rotating objects. I also
adopt the assumption that charged particle beams extracted from the rotating object by
electric fields compose these currents; see Arons (2002) for a recent discussion of these currents'
significance, and a discussion (briefly recapitulated below) for the evidence that such
beams actually exist. 

If conduction currents are the sole source of the electromagnetic fields, and if 
a charged particle beam carries all of the current,
$\dot{E}_R = I \Phi_{\rm mag} = qc \dot{N} \Phi_{\rm mag}$, and
\begin{eqnarray}
\dot{N} & = &  c \left[\frac{\Omega B_{*,{\rm dipole}}}{2\pi c} \right] 
   (2 \pi A_{\rm cap}), \; \; \; 
\label{eq:particle_loss} \\
& = &  \frac{\Omega^2 \mu}{|q|c} = \frac{\dot{E}_R}{|q| \Phi_{\rm mag}}
\label{eq:GJloss}
\end{eqnarray}
One recognizes $A_{\rm cap} \pi (\Omega R_* /c) R_*^2$ to be the area of a magnetic polar 
cap, if the closed field lines extend to the light cylinder, and the expression 
in square brackets in (\ref{eq:particle_loss}) to be the Goldreich-Julian charge 
density $\rho_{GJ} $ at the stellar surface, the number of elementary charges 
per unit volume such that parallel  electric 
fields are shorted out in the magnetosphere. 

The electric current 
\begin{equation}
I_{GJ} = 2 A_{cap} c \rho_{GJ} = \frac{\Omega_*^2 \mu}{c} = c \Phi_{\rm mag}
\label{eq:GJ_current}
\end{equation}
is a fundamental item in the theory of the spindown of relativistic magnetized 
rotators. Since $\dot{E}_{EM}= I_{GJ} \Phi_{\rm mag} $, rotators which have 
such currents linking the star to the outside world can spin down at a 
rate comparable to that expected from vacuum (magnetic dipole) radiation 
losses. Both contributions to spindown are expected to be comparable, in
oblique rotators - there is substantial evidence,
from the lack of dependence of radio pulsars' spindown torques on the angle between the
rotation and dipole axes (Lyne and Manchester 1988) that this assumption is true. 

Such currents, in stars with dipolar magnetic fields, include a primary
charged particle beam emanating from the polar cap - electrons, when
$i \equiv \angle ({\bf \mu, \Omega}) < 90^\circ $. That current must be balanced, at least on 
average, by an electric return current. The most traditional ideas assume 
this return current is an outflowing beam of charges of the opposite sign - ions, if
$i < 90^\circ$ - usually 
assumed to occur on the ring of field lines around the polar cap which form 
the boundary layer separating the closed and open field line zones of the 
magnetosphere (Goldreich and Julian 1969); in force free models (Michel 1975, 
Contopoulos {\it et al.} 1999), the return current appears as a current sheet. Such ``auroral'' field 
lines map to the rotational equator, in oblique as well as aligned rotators, where the return current
forms a (corrugated) current sheet, with spatial oscillations imposed by the
rotation of the underlying non-aligned dipole magnetic field. The model developed here for 
UHECR assumes the ion injection rate per magnetar to be that of this electrodynamically fundamental 
{\it return} current, based on an analogy to what has been learned from Pulsar
Wind Nebulae.  

$\dot{E}_{EM}= I_{GJ} \Phi_{open} $ does {\it not}
mean that such an object automatically is a maximal particle accelerator, with all the
energy going into the energy of the particles in the electric currents. For that to be true,
the particles in the currents would have to transport all of the rotational
energy lost, and each would have have to reach the maximum energy per particle $q\Phi_{\rm mag}$ 
- other particle constituents would have to be
largely absent, and Poynting fluxes would have to be unimportant.  Pulsar theory,
and related theories of magnetized disks,  
begins with the opposite
assumption - that a large scale Poynting flux carries almost all of the 
outlowing energy, with the current carriers gaining very little of the potential 
energy available until
dissipation occurs in the region where the outflow from the star 
terminates (Goldreich and Julian 1969). Even then, 
the energy dissipated might go
into bulk expansion of the surroundings, rather than into acceleration of a few
high energy particles. This approximation - represented in its most extreme form
by force free models of the outflow ({\it e.g.}, Contopoulos {\it et al.} 1999) - gets its 
emprirical support from the lack of 
intense photon emission from many known relativistic outflows. Therefore, one cannot
blithely assume the existence of an
energy/particle given by (\ref{eq:particle-energy}) with $\eta \sim 1$, even when the ion 
loss rate is given by (\ref{eq:GJloss}).

In addition, the study of 
pair creation in the magnetosphere has
led to the belief that a dense  electron-positron plasma provides the 
dominant constituent by number in the outflow.  
The feeding of the X-ray source in the 
Crab Nebula, in the X-ray nebula around the Vela pulsar (Helfand {\it et 
al.} 2001) and in the nebula G320.4 around PSR 1509-58 (Gaensler {\it et al.} 2002) 
strongly supports the existence of such a dense plasma 
outflow - if the flow leaving the neutron star has the same composition on 
all field lines, the particle input to the Crab 
($\dot{N}_{\rm total} \geq 10^{38}$ pairs/second) suggests $\eta_{pairs} < 10^{-3}$ 
- the pairs carry much of the energy flow at large distances from 
the star, but being a dense plasma, they have energy per particle well less 
than (\ref{eq:particle-energy}). 

However, study of the structure of the shock wave terminating the
equatorial relativistic wind in the Crab ({\it e.g.} Hester {\it et 
al.} 1995), in G320.4 (Gaensler {\it et al.} 2002) and perhaps in the Vela
nebula (Helfand {\it et al.} 2001), where the energy in the 
pulsar wind gets transfered to the plasma forming the 
observed synchrotron nebulae, suggests that an interestingly large fraction of the total
voltage does get applied to a Goldreich-Julian flux of
heavy ions on the equatorial flow lines from rotation powered pulsars (Arons 2002). 
These are probably light nuclei 
such as protons or alpha particles (Gallant and Arons 1994). Detailed modeling of this 
interaction zone (Gallant and Arons 1994,  Spitkovsky and Arons 2000 
and submitted to ApJ) suggests that the total ion injection rate is indeed the Goldreich-Julian
rate (\ref{eq:GJloss}) and that these these particles, flowing out in a
latitudinal sector around the stars' rotational equators, 
{\it have} experienced $\sim $ 10\% of the
full potential drop ({\it i.e.}, $\eta \sim 0.1$), and thus carry an energy loss
from the rotator 
$\dot{E}_{ions} = I_{GJ}\eta \Phi_{mag} = \eta \dot{E}_R \sim 0.1 \dot{E}_R $ of
each neutron star's rotational energy loss in an equatorial sector
filling $\sim 1/5$ of the sky around each of these pulsars, exactly the region 
where the models hypothesize the flow of the return current in an equatorial current sheet. 
Outside this equatorial sector, the wind is
more likely to be Poynting flux (AC and DC) dominated.   

The specifics of the ion accelerator are not 
known - most recent suggestions for the acceleration site have focused 
on non-ideal processes in the equatorial wind, well outside the light cylinder. The 
ions enter the wind, having been drawn up from the star along the boundary 
layer separating the closed and open field lines, leave the star in the plane of the star's 
rotational equator,  and gain their high energy per particle at some large 
distance from the star, perhaps at the interesting radius where the 
shortage of charge carriers causes dissipation of the corrugations in the equatorial current sheet 
(Michel 1971, Coroniti 1990, Michel 1994), or transform to electromagnetic
waves, either Alfvenic (Bellan 1999, 2001) or vacuum-like (Melatos and Melrose 1996; Melatos 1998),
whose pomderomotive force can accelerate the particles, through surf riding in the EM fields,
 to energies comparable to $q\Phi_{mag}$. 
I discuss surf riding acceleration
mechanism further in \S \ref{sec:accel}, but a detailed study of that physics
is outside the scope of the present investigation.

In what follows, I assume that such ion acceleration also occurs in 
magnetars that were rapidly rotating at birth, and explore the
consequences of this assumption for an extragalactic origin of UHECRs
from {\it normal} galaxies. I assume all galaxies which have core collapse
supernovae form magnetars with a birthrate $\nu_m$
comparable to that in our own galaxy, where $10^{-3} > \nu_m > 10^{-5} $ 
years$^{-1}$ (Gaensler {\it et al.} 2001), and that some are born with
high angular velocities, $\Omega_{i} \sim 10^4 $ s$^{-1}$. 
During their spindown, they emit  heavy ions (taken to be 
protons in what follows) with an ion loss rate per magnetar
given by (\ref{eq:GJloss}), with the energy/particle in the emitted 
equatorial ``beam'' declining as the rotator ages, as specified by
(\ref{eq:particle-energy}), with $\eta \approx 0.1$ in analogy to what has been
inferred for pulsars. Therefore, the highest 
energy particles come from new born objects. The particles cannot be 
contained within normal galaxies, therefore the flux at the earth reflects the 
contributions from all galaxies within the energy dependent GZK volume. 
That volume expands to encompass all galaxies in the 
Hubble volume, as the particle energy decreases, but below the ankle energy $E_a$, 
this extragalactic source becomes swamped by the softer particle spectrum thought to be
due to acceleration in our own galaxy.

A simple estimate shows that the idea is a viable candidate for the origin 
of the UHE spectrum. The star emits ions at the rate
(\ref{eq:GJloss}).  Suppose electromagnetic torques are the sole means of 
spinning down the neutron star.  The strong $\Omega $ dependence says 
that the star emits most of the
particles during the initial loss of rotational energy, when 
$\Omega = \Omega_i = 10^4 \Omega_4 $.  The 
initial electromagnetic spindown time is
\begin{equation}
\tau_{EM} = \frac{\frac{1}{2}I \Omega_i^2}{\dot{E}_{EM}}
   =\frac{9}{8} \frac{Ic^3}{\mu^2 \Omega_i^2}
     = 5 \: \frac{I_{45}}{\mu_{33}^2 \Omega_4^2} \; {\rm minutes}.
\label{eq:tau_EM}
\end{equation}
Therefore, each magnetar injects 
\begin{equation}
N_i \sim \dot{N}_{GJ}(\Omega_i) \tau_{EM}(\Omega_i)  = \frac{Ic^2}{Ze\mu} 
    = 4 \times 10^{42}  \frac{I_{45}}{Z \mu_{33}}
\label{eq:initial-burst}
\end{equation}
ions into the metagalaxy, each with the energy (\ref{eq:particle-energy}).  
The existing observations of a few events at energies $> 10^{20}$
eV require the birth of some magnetars with $ \eta_1 \Omega_4^2 \mu_{33} > 0.1/Z$.

With a magnetar birth rate per galaxy $\nu_m = 10^{-4} \nu_{m4} \; {\rm yr}^{-1}$ 
and a galaxy density 
$n_g = 0.02 n_{g2}$ Mpc$^{-3}$, the number of particles per unit volume
from magnetars in all the galaxies is
$
n_{UHE, magnetars}^{(est)} = N_i n_g \nu_m T_{loss}(E_a),
$
where $T_{loss} (E_a) $ is the residence time of a particle at the ankle initially
injected at higher energy; very roughly, this is the lifetime $T_\pm (E_a)  \approx 10$ Gyr. 
Then 
\begin{equation}
J^{(est)}(>E_a) \sim \frac{c}{4\pi} N_i \nu_m n_g T_\pm (E_a)  \approx 
     7 \times 10^{-18} \left(\frac{I_{45}}{Z \mu_{33}}\right) \nu_{m4} 
       \; {\rm cm^{-2}-sec^{-1}-ster^{-1}},
\end{equation}
comfortably above the observed $J(>E_a) = 3 \times 10^{-18} \:
{\rm cm^{-2} - sec^{-1} - ster^{-1}}$. 

The required luminosity in UHE particles does not unduly tax
the energetics of rapidly rotating magnetars in galaxies, as
can be seen by writing $L_g^{UHE} = \epsilon_m \nu_m E_R^{(i)}$.  With $L_g^{UHE}$
from (\ref{eq:UHE_gal_lum})  and
$E_R^{(i)} = (1/2) I \Omega_i^2 = 5 \times 10^{52} I_{45} \Omega_4^2 $ ergs,
the data require a fraction
$\epsilon_m = 0.005/\nu_{m4} n_{g2} I_{45} \Omega_4^2 $
of the initial rotational energy to go into UHE particle acceleration. I show in \S \ref{sec:supershells}, 
using the results of the more detailed theory in \S \ref{sec:spectrum}, that the birth rate of
fast magnetars, those with initial voltages large enough to accelerate particles with energy
greater than $E_a$, is  $\nu_m^{(fast)} = (0.05 - 0.1) \nu_m$.  Then the coefficient in 
$\epsilon_m$ changes from 0.005 to
.05-0.1, still not a taxing demand on the model's energetics. Put slightly differently,
the luminosity density in rapidly rotating magnetars deduced from the fit of the model
to the observations described in \S \ref{sec:spectrum} is
$\sim \nu_m^{(fast)} n_g E_R^{(i)} = 10^{46} (\nu_m^{(fast)}/10^{-5} \: {\rm yr}^{-1})
 I_{45} \Omega_4^2 \; {\rm ergs/Mpc^3-year}$.  Given that 10\% of the accelerating voltage
gets applied to each accelerated particle, and that the injection rate per object is given
by (\ref{eq:GJloss}), the model has 
the required luminosity density $\sim 10^{45}$ ergs/Mpc$^3$-year in UHECRs.

As I shall show, more careful
treatment of the spectrum, plus approximate accounting for the losses
sustained as the magnetized wind bursts out of the supernova that gives
rise to the neutron star, improves the correspondence between the model and
the data, and makes some interesting predictions for future UHECR observations.

\section{The Spectrum of UHE Cosmic Rays from Bare Metagalactic Magnetars \label{sec:spectrum}}

\subsection{The Injection Spectrum from an Isolated Magnetar \label{sec:injection}}

During a time $dt$, the magnetar spins down from angular velocity 
$\Omega(t)$ to $\Omega(t+dt) = \Omega(t) + \dot{\Omega} dt$.
According to (\ref{eq:particle-energy}) and (\ref{eq:GJloss}), the energy/particle 
$E(t) = m_ic^2 \gamma (t) $ and the instantaneous injection rate $\dot{N_i}$ each decay 
$\propto \Omega^2$. Therefore,
the number of particles accelerated as the star spins down in time $dt$
with energy between $mc^2 \gamma $ and $mc^2 (\gamma - d\gamma)$ is
\begin{equation}
-dN_i = \dot{N}_i [\Omega(t)]dt = \dot{N}_i 
   \frac{dt}{d\Omega} \frac{d\Omega}{d\gamma} d\gamma,
\label{eq:dNi}
\end{equation}
with
\begin{equation}
\frac{d\Omega}{dt} \equiv \dot{\Omega} = -\frac{\dot{E}_R}{I\Omega} = 
 - \frac{\dot{E}_{EM} + \dot{E}_{grav}}{I \Omega}.
\label{eq:spindown}
\end{equation}
Here $I$ is the moment of inertia, and (\ref{eq:EMloss}) expresses
the electromagnetic energy losses. 

The gravitational wave losses may be substantial.  If the star has
a static equatorial ellipticity $\epsilon$, gravitational waves extract 
energy at the rate (Ostriker and Gunn 1969)
\begin{equation}
-\dot{E}_{grav} = \frac{32}{5} \frac{GI^2 \epsilon^2 \Omega^6}{c^5}
 = 1.8 \times 10^{51} I_{45}^2 \Omega_4^6 \epsilon_2^2 \; {\rm ergs/sec},
\label{eq:static_GR}
\end{equation}
where $I_{45} = I/10^{45}$ g-cm$^2$ and $\epsilon_2 = \epsilon/10^{-2}$.
Note that $\epsilon_2 \sim 1 $ is a substantial but not ridiculous ellipticity; 
interior magnetic fields can create ellipticity 
(Usov 1992, Bonnazola and Gourgoulhon 1996) with magnitude
(Ostriker and Gunn 1969)
\begin{equation}
\epsilon = \frac{25}{24} \frac{R^4}{GM^2} 
     (3B_{poloidal}^2 - \langle B_{toroidal}^2 \rangle) 
\approx 10^{-2} \frac{3B_{poloidal}^2 - \langle B_{toroidal}^2 \rangle}
      {(4 \times 10^{16} \; {\rm Gauss})^2};
\label{eq:mag_ellipt}
\end{equation}
the numerical value assumes $M = 1.4 M_\odot$.
The initial spindown time due to gravitational radiation is 
\begin{equation}
\tau_{GW} = \frac{\frac{1}{2}I\Omega_i^2}{-\dot{E}_{grav}} 
     = \frac{28.5}{I_{45} \epsilon_2^2 \Omega_4^4} \; {\rm seconds}.
\label{eq:tau_GW}
\end{equation}

Interior fields an order of magnitude stronger than the surface field are by no
means incredible. Formation of interior fields as large as $10^{17.5}$ Gauss
in magnetars has been proposed, as 
a consequence of dynamo activity 
in the first few seconds of the proto-neutron star's life (Duncan and  Thompson 
1992, Wheeler {\it et al.} 2000.) - these stars are thought to have
non-dipolar surface fields much stronger than the already large dipole field.
Once formed, such an ellipticity 
would most likely persist throughout the neutron star's spindown, certainly through the
first day or so of the spindown germane to the model discussed here.

Other sources of ellipticity or its equivalent may be dynamical instability 
of the star, if the
rotational energy exceeds 25\% of the newly formed star's gravitational energy,
and secular instabilites of F- and R-modes, driven by the gravitational radiation 
itself (e.g., Lai 2001, Ushomirsky 2001).
In the interest of 
simplicity, I use (\ref{eq:mag_ellipt}) as motivation to choose 
$\epsilon_2 \sim 1$ as a parameter throughout the subsequent discussion, 
and neglect other versions of gravitational wave spindown. In the appendix,
I briefly summarize the possible effects of
gravity waves emitted because of the R mode instability. 

I rewrite the spin down rate as
\begin{equation}
-\dot{\Omega} = \frac{4}{9} \frac{\mu^2 \Omega^3}{I c^3}
       \left[1 +\left(\frac{\Omega}{\Omega_g}\right)^2 \right]
\label{eq:spindown2}
\end{equation}
with $\Omega_g$ the angular velocity at which gravity wave and electromagnetic 
losses are equal,
\begin{equation} 
\Omega_g \equiv \left(\frac{5}{72} \frac{c^2 \mu^2}{GI^2 \epsilon^2}
        \right)^{1/2} 
       = 3 \times 10^3 \frac{\mu_{33}}{I_{45} \epsilon_2} 
          \; {\rm s}^{-1}.
\label{eq:omega_g}
\end{equation}
Expressions (\ref{eq:particle-energy}), (\ref{eq:GJloss}), (\ref{eq:dNi}) and 
(\ref{eq:spindown2}) then yield the particle spectrum accelerated
by a magnetar during its spindown to be  
\begin{equation}
\frac{d N_i}{d\gamma} = 
  \frac{d N_i}{dt}\left(-\frac{dt}{d\Omega}\right)mc^2 \frac{d\Omega}{dE} =
  \frac{9}{4} \frac{c^2 I}{Z e \mu \gamma}
       \left( 1 + \frac{\gamma}{\gamma_g} \right)^{-1},
\label{eq:diff-inj-rate}
\end{equation}
where
\begin{equation}
E_g = m_ic^2 \gamma_g  =  \frac{Z \eta e \mu}{c^2} \Omega_g^2 = 
               \frac{5}{72} \frac{Z \eta e \mu^3}{G I^2 \epsilon^2}
   =  3 \times 10^{20} \frac{Z \eta_1 \mu_{33}^3}{I_{45}^2 \epsilon_2^2}
                \; {\rm eV}. 
\label{eq:crit_grav_E}
\end{equation}

The star promptly emits the energy that goes into the wind. The time to 
spin down to a voltage such that a particle of energy $ E $ 
can be accelerated is 
\begin{eqnarray}
\frac{t_{spin}(E)}{\tau_g} & = & \left(\frac{\Omega_g}{\Omega} \right)^2 -
       \left(\frac{\Omega_g}{\Omega_i} \right)^2 -
        \ln \left[\frac{\Omega_i^2}{\Omega^2} 
       \frac{1+ \frac{\Omega^2}{\Omega_g^2}}{1+ \frac{\Omega_i^2}{\Omega_g^2}} 
         \right] \label{eq:spin_hist} \\
     &= & \frac{E_g}{E}  - \frac{E_g}{E_i}  -
        \ln \left(\frac{E_i}{E} 
       \frac{1+ \frac{E}{E_g}}{1+ \frac{E_i}{E_g}} 
         \right)
\label{eq:time_to_E},
\end{eqnarray}
where
\begin{equation}
\tau_g = \frac{\frac{1}{2}I\Omega_g^2}{\frac{4}{9} \frac{\mu^2 \Omega_g^4}{c^3}}
       = \frac{81}{5} \frac{G I^3 c \epsilon^2}{\mu^4}
       = 0.9 \frac{I_{45}^3 \epsilon_2^2}{\mu_{33}^4} \; {\rm hours}
\label{eq:tau_g}
\end{equation}
is the electromagnetic spin-down time, for a star spinning at angular frequency
$\Omega_g$. (\ref{eq:tau_EM}) and (\ref{eq:tau_GW}) give the  
initial spin-down times for pure electromagnetic and pure gravity wave losses.
Expression (\ref{eq:spin_hist})
yields the spindown history when gravity wave losses dominate,
$\Omega(t) = \Omega_i [1+(2t/\tau_{GW})]^{-1/4}$, if the limit $\Omega_g \ll \Omega < \Omega_i$
exists, and yields pure electromagnetic
spindown, $\Omega(t) = \Omega_i [1+(t/\tau_{EM})]^{-1/2}$, in the limit
$\Omega \ll \Omega_g$. For $t < 2\tau_g$, 
(\ref{eq:time_to_E}) simplifies to
\begin{equation}
t_{spin}(E) =  2 \tau_{GW} \left[ \left( \frac{\Omega_i}{\Omega} \right)^4 -1 \right]
  \approx 1.8 \frac{Z^2 \eta_1^2 \mu_{33}^2}{I_{45} \epsilon_2^2} 
      \left(\frac{10^{20.5} \, {\rm eV}}{E} \right)^2 \: {\rm hours}, \: 
       E_g < E < E_{max}.
\label{eq:gr_time_to_E}
\end{equation}
Here $E_{max} = Ze\eta \Phi_{\rm mag}(\Omega_i) = 3.3 \times 10^{21} Z\eta_1 \mu_{33} \Omega_4^2 $
eV. For $t> 2\tau_g, \; E < E_g $, electromagnetic losses alone accurately describe the spindown.  
In this limit,  
\begin{equation}
t_{spin} (E) = 2 \tau_{EM} \left( \frac{\Omega_i^2}{\Omega^2} -1 \right) \approx
     1.8 \frac{Z\eta_1 I_{45}}{\mu_{33}^2} \frac{10^{20.5} \; {\rm eV}}{E} \: {\rm hours},
     \; E < E_g.
\label{eq:em_time_to_E}
\end{equation}
The general spin history can be written in the simplified form, valid for
$t \gg \tau_{GW}$,
\begin{equation}
\Omega = \Omega_g \left( \frac{2 \tau_g}{t} \right)^p,
\label{eq:simple_spin_hist}
\end{equation}
with $p = 1/4$ for $\tau_{GW} \ll t < 2\tau_g$ and $p = 1/2$ for $t > 2\tau_g $.

Expression (\ref{eq:diff-inj-rate}) is the spectrum that would be observed
from a single nearby event by an experiment which accumulates events of energy $E$ and higher,
over the time (\ref{eq:time_to_E}). An experiment which resolves a single 
nearby magnetar birth and spindown
would see an instantaneous, approximately monoenergetic spectrum, with the energy/particle
declining in proportion to $\Omega^2 (t)$.

To the magnetar's host 
galaxy and to the metagalaxy, the energy input to the
relativistic wind  appears as an impulsive burst; for the energies of all the observed
particles ($E_{max} \gg E > E_a = 6 \times 10^{18}$ eV), 
$t_{spin} (E) < t_{spin}(E_a) = 42 Z\eta_1 I_{45} /\mu_{33}$ hours.
Note that this is the time taken to put energy {\it into} the wind, corresponding
to voltages high enough to accelerate the observed ultra-high energy cosmic rays;
the actual time of acceleration is later. How much later depends on the time it takes for
a region of the wind to flow to the radius where acceleration occurs. In \S \ref{sec:accel},
I suggest this flow time probably is a few hours, for typical parameters.

\subsection{Intergalactic Particle Transport and the Observed Spectrum from Bare Magentars}

\subsubsection{Scatter Free Transport}

The magnetars forming in galaxies emit UHE particles which cannot
be contained within the galaxies.  Therefore, normal galaxies inject
particles into intergalactic space at the average rate per unit volume
$n_g \nu_m dN_i /d\gamma $. Assuming scattering in 
intergalactic magnetic fields has negligible effect on particle transport,
the intergalactic spectrum $N(\gamma) $ may be determined from
\begin{equation}
\frac{\partial}{\partial \gamma} 
   ( \dot{\gamma}  N) =
     W_{geom} \nu_m n_g \frac{dN_i}{d\gamma}.
\label{eq:continuity_eq}
\end{equation}
Here $\dot{\gamma} = - \gamma /T_{loss}(\gamma)$ and $W_{geom} = 0.5 $ represents 
the geometric fact that only half the magnetars 
are expected to have the magnetic geometry appropriate
to injecting ions from the stars atmospheres into the wind in the rotational equator
(\S \ref{sec:MWNescape}).

Integrating equation (\ref{eq:continuity_eq}) with (\ref{eq:loss-rate})
and  converting from
$N(\gamma) $ to $J(E)$ through $m_i c^2 J = c N/4 \pi $ yields
\begin{equation}
J(E) = \frac{c}{4\pi m_i c^2} N(\gamma) =
W_{geom}  \frac{9}{16\pi} \frac{Ic^3}{Z e \mu} 
 \frac{ n_g \nu_m T_{loss}(E) }{E } 
\ln \left(\frac{E_{max}}{E}\frac{1+\frac{E}{E_g}}
  {1+\frac{E_{max}}{E_g}}\right).
\label{eq:spectrum}
\end{equation}

The spectral form has the $1/E $ shape of 
the injection spectrum below
10 EeV, steepens to become proportional to $E^{-3}$ in the {\it observed} UHE
range $10 \; {\rm EeV} < E < 100 \; {\rm EeV} $, then flattens to return to 
a shape approaching the  $E^{-1} $ spectrum above 300 EeV 
($\propto E^{-2}$ if gravitational wave losses during spin down truncate 
the number of particles produced at energies well below $E_{max}$, {\it i.e.},
when $E_g \ll E \leq E_{max}$),
where GZK losses saturate. In the strong GR case, gravity wave emission shortens the
time at which each magnetar can contribute particles at a given energy, which reduces 
the flux even in the better studied region $ E < 5 \times 10^{19}$ eV, for 
parameters other than ellipticity fixed.  

These ultra high energy particles have little trouble entering our galaxy\footnote{The 
ion Larmor radius 
$r_L = E/ZeB = 2.3 (E/10^{18.8} \: {\rm eV}) (3 \: \mu{\rm Gauss}/ZB_0 \exp(z/2h)$ 
kpc, where $z$ is the vertical distance from the galactic midplane and $B_0$ is the midplane magnetic field, 
is larger than the vertical extent of the synchrotron halo of our galaxy, whose scale height
is $h \sim $ 1.8 kpc (Beuermann {\it et al.} 1985) implying easy entry even at the low end of
the UHE spectrum. An extended, quasi-spherical magnetized halo, if it exists, probably has a magnetic 
field at least an order of magnitude weaker (with Larmor radius $\sim $ 20 kpc), 
an assumption supported by radio synchrotron emission
observed in other disk galaxies (e.g, Hummel {\it et al.} 1991, Lisenfeld and V\"{o}lk
2000). Thus even at the lowest observed UHECR energies, individual particles can reach our galaxy's midplane
following almost rectilinear orbits, in times not much exceding the light transit time - at the lowest energies,
the guiding center drift velocity might drop a bit below c, to $\sim c (r_L/R_{halo})$, if an extended halo
with radius larger than 20 kpc exists.  
Since energy losses are negligible on this
scale (and they would be negligible, even if the particles had to diffuse in the galaxy's magnetic field), 
the particle density and spectrum inside our galaxy comes to equilibrium with the intergalactic spectrum
at the same density and spectrum as is in the metagalaxy (Ginzburg and Syrovatskii 1964, Longair 1994). 
However, magnetic deflection in halo magnetic fields probably does corrupt
the accuracy with which one can infer a possible source's location from the look back direction
along an air shower (itself known to $\sim 1^\circ $ accuracy) for energies below $5 \times 10^{19}$ eV.}. 
Therefore, expression (\ref{eq:spectrum}) can be compared directly to the observations.
Multiplying (\ref{eq:spectrum}) by $E^3$ and overlaying the results on the data
yields the result shown in Figure 1.  Given the substantial
errors, fitting by eye suffices,
with the conclusion that this elementary theory provides a satisfactory fit to the
data for the values of $K_0 = W_{geom} \nu_m I n_g T_H/\mu $ shown in 
Table \ref{tab:source_strength}.

\begin{figure}[H]
\unitlength = 0.0011\textwidth
\begin{center}
\begin{picture}(950,950)(0,15)
\put(0,100){\makebox(950,950){\epsfxsize=950\unitlength \epsfysize=950\unitlength
\epsffile{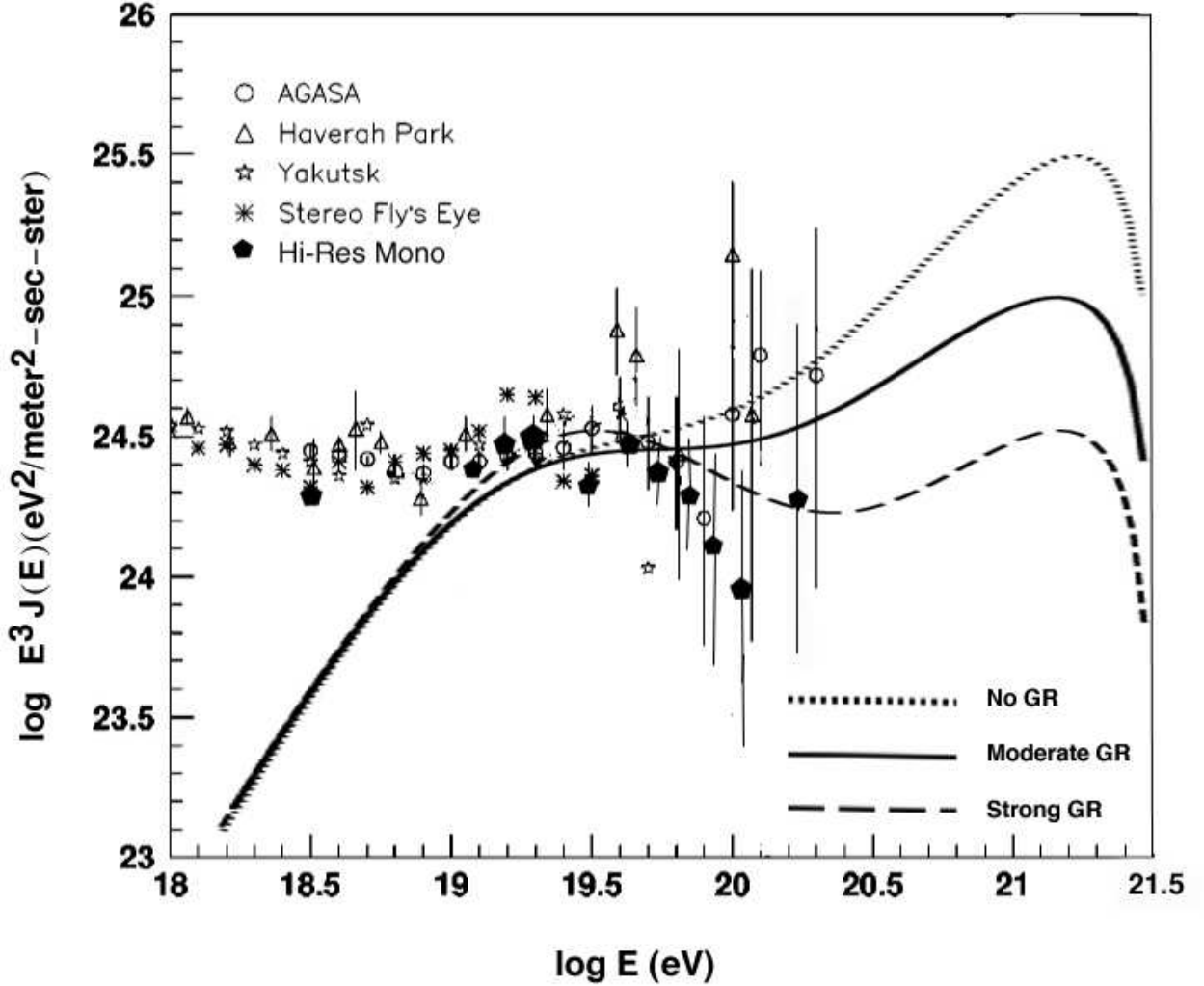}}}
\end{picture}
\hspace{1\unitlength}
\vspace{-100\unitlength}
\caption{\label{fig:theory_data} Comparison of bare magnetar theory to UHECR data. 
The experimental data are
from Nagano and Watson's (2000) summary except for the data from the Hi-Res (monocular)
results, which are taken from Abu-Zayyad {\it et al.} (2002b). The three theoretical curves
are for differing assumed equatorial ellipticities of the neutron stars: $\epsilon = 0$
(``no GR''); $\epsilon = 10^{-2}$ (``moderate GR''); $\epsilon =10^{-1}$ (``strong GR'').
The theoretical curves were fit to the data by eye, by requiring the model flux at 
$3 \times 10^{19}$ eV to provide the whole intensity at this energy. That fit
yields values for the average source strength 
$K_0 = W_{geom} \nu_m I n_g T_H/\mu $, which are listed for the three cases shown in Table
\ref{tab:source_strength}. All the 
curves use the same upper cutoff energy $E_{max} = 3.3 \times 10^{21}$ eV, corresponding
to $Z \eta_1 \mu_{33} \Omega_4^2 =1 $ (a polar dipole field strength of $2 \times 10^{15}$  
Gauss and an initial rotation period of 0.6 msec.) Lowering the upper cutoff to the minimum
acceptable value of $3 \times 10^{20}$ eV yields qualitatively unacceptable fits to the 
observations - there is too much curvature in $E^3 J(E)$ below $6 \times 10^{19}$ eV. 
Including a ``galactic'' component 
$E^3 J_{galactic}(E) = 7 \times 10^{23} (E/ 30 \; {\rm EeV})^{-3.2} $ 
ev$^2$/meter$^2$-sec-ster constructed to represent the data at
energies well below $E_a$ and extrapolated to {\it all} energies above $E_a$ 
reduces the inferred values of $K_0$ by a factor 0.65.
These modified values are also shown in Table \ref{tab:source_strength}.}
\end{center}
\end{figure}

\placefigure{fig:theory_data}

\setcounter{figure}{1}
\noindent At the relatively low energies now
observed, the presence or absence of gravitational wave losses makes little difference unless
the gravity wave losses are large.
Below 4 EeV, this extragalactic particle spectrum hides below the flux of
(probably)  galactic particles at energies below the ankle.

The spectrum in Figure 1 represents an average over many particle bursts, each lasting 
a couple of days (for $E > E_a$). These bursts should be distinguishable as such, with
sufficient collecting area. One can readily show, using the time dependent form of
(\ref{eq:continuity_eq}) with its right hand side set to zero and with (\ref{eq:diff-inj-rate})
as the initial condition, that the fluence ${\cal F}$ of a single burst has the same spectral form as
is shown in Figure 1, but with the upper cutoff energy now determined by the losses in
the CMB degrading the maximum injection energy to the value determined by the source's distance.

Because the injection spectrum is extremely flat ($\propto E^{-1}$, if gravitational wave losses
are negligible), the losses in the CMB flatten $E^3 J$ but do not produce the GZK pileup
characteristic of steeper injection spectra ({\it e.g.}, Berezinsky {\it et al.} 2002.) If the claimed 
observation of such pile up by the Hi-Res experiment (Abu-Zayyad {\it et al.} 2002 a,b) is confirmed
and the discrepancy with the AGASA results (Hayashida {\it et al.} 1994, Takeda {\it et al.}
1999; see http://www-akeno.icrr.u-tokyo.ac.jp/AGASA/ for results through Fall 2001) is resolved, 
the existence of such a feature would,
within the context of the model desrcibed here, point to very rapid early spindown due to gravitational
wave losses, which has the effect of steepening the highest energy part of the injection spectrum -
see the ``strong GR'' curve in Figure 1.

\placetable{tab:source_strength}

\begin{table}[H]
\begin{center}
\begin{tabular}{lcccc}
\hspace*{1in} & $\epsilon$ & $E_g$ (eV) & $K_0$ & $K_0^{(eff)}$ \\ \hline
no GR & $ < 0.003 $  & $\infty $ & 0.03 & 0.02\\
moderate GR & 0.01 &  $3 \times 10^{20}$ & 0.06 & 0.04\\
strong GR  & 0.1 & $3 \times 10^{18}$ & 1.25 & 0.81
\end{tabular}
\end{center}
\caption{Source strength 
  $K_0 = W_{geom} (\nu_m /10^{-4} \; {\rm yr}^{-1})(n_g /0.02 \; {\rm Mpc}^{-3} )
  (I_{45} / \mu_{33}) (T_H / 14 \; {\rm Gyr})$ for the theoretical curves shown in 
  Figure 1. $K_0^{(eff)}$ is the source strength inferred if one extrapolates
  the spectrum found at energies well below $E_a$ into the UHE regime, then
  superposes the model.\label{tab:source_strength}}
\end{table}

\subsection{Predicted Spectrum at Higher Energies \label{sec:predicted}}

Unless current experiments are so fortunate as to have observed the upper cutoff, 
the spectrum of high energy particles
should continue into the higher energy region where current detectors lack
sensitivity. The specific 
model used here suggests the spectrum $E^3 J(E)$ should resume rising with $E$ above the
energy $E_G = 2.8 \times 10^{20}$ eV where the GZK loss rate becomes 
approximately energy independent (unless the gravitational wave losses
are large, as shown in Figure 1.) 

This prediction 
is simplified, in its assumption that
all magnetars have exactly the same starting voltage ($10^{22.5}$ Volts) and single 
particle efficiency  $\eta \sim 0.1$.  The inferred value of $K_0$ in Table 1 being small compared 
to unity in the weak and 
moderate GR cases has its most natural interpretation in the initial voltages (initial rotation
periods, if all objects have the same dipole moment) having a distribution of values, with only
a small fraction having voltages such that particles of energy greater than $E_a$ can be
accelerated. Since the results of \S \ref{sec:MWNescape} yield $W_{geom} = 0.5$, 
Table \ref{tab:source_strength} suggests that the birth rate of magnetars with voltages high enough
to contribute to the spectrum in Figure 1 is only 6\% of the overall
magnetar birth rate, in the ``no GR'' case (a ``fast'' magnetar formation
rate of once every 170,000 years/galaxy); for
``moderate GR'', the required rate is 10\% of the overall magnetar birth rate (once every $10^5$
years/galaxy). 

A distribution of initial voltages can have potentially observable effects on the predicted
UHECR spectrum.  To illustrate, suppose the distribution of birth rates as a function of starting voltage
is a power law,
\begin{equation}
\frac{d\nu_m}{d\Phi_i} = \frac{\nu_m}{\Phi_{i,max}} 
      \frac{s-1} {\left(\frac{\Phi_{i,max}}{\Phi_{min}} \right)^{s-1} - 1 }
      \left( \frac{\Phi_i}{\Phi_{i,max}} \right)^{-s} ,
\label{eq:birth_spectrum}
\end{equation}
with $\Phi_{min} \leq \Phi_i \leq \Phi_{i,max}$, 
$\Phi_{i,max} = 3 \times 10^{22} \mu_{33} (\Omega_i /10^4 \; {\rm sec}^{-1})^2 $
and $\Phi_{min} = 10^{14} $ Volts, the potential drop characteristic of the known
magnetars - as with pulsars, initial rotation periods comparable to the observed
periods are possible, so that initial rotation periods as long as 10 seconds must
be considered. Expression (\ref{eq:spectrum}), with $d\nu_m /d\Phi_i$ from ({\ref{eq:birth_spectrum}) 
replacing $\nu_m$,
now gives $E^3 \partial J(E; \Phi_i)/\partial \Phi_i$, with $\gamma_{max} $ replaced by
$\gamma_i \equiv Ze\eta \Phi_i /m c^2$. Integrating over all initial voltages has the effect,
when $s > 1$,
of replacing the logarithmic cutoff in (\ref{eq:spectrum}) by
$(E /E_{i,max})^{-(s-1)} S(E/E_{i,max}, s)$, where $S$ is a slowly varying function which
produces a more complicated logarithmic cutoff of the spectrum at 
$E_{i,max} \equiv Ze\eta_1 \Phi_{i,max}$. If $s < 1$, there is no substantial change from
the single initial voltage model - when the spectrum of birth potentials is flat ($s <1$),
essentially all the stars would have starting voltage
equal to $\Phi_{i,max}$. 

Fitting the resulting modified spectrum to the
data (again by eye)
yields the required reduction of the normalization if $s \approx 1.2 $, in the ``no GR'' case. The
extra $E^{-0.2}$ now introduced in $E^3 J(E)$ at energies below $10^{20}$ eV is 
undetectable in the current experiments, but with increased data above $5 \times 10^{19}$
eV the shape of the spectrum will provide significant constraints on such refinements
of the model. Above $10^{20}$ eV, the decreased number of contributing sources can substantially
reduce the high peaks at high energy shown in Figure 1, by as much as
a factor of 3. However, the basic conclusion, that the model predicts something to see, roughly
at the level one would guess by extrapolating the observed $E^3 J(E)$ above the formal GZK cutoff, 
remains robust. With $s = 1.2$, 5\% of all the magnetars have $Ze\eta \Phi_i$ above $E_a$, 
consistent with the simple inference of the fast magnetar birthrate described above.

Also, the highest energy part of the spectrum probes acceleration in
the wind at the earliest times.  As discussed in \S \ref{sec:escape}, breakout of the wind
from a magnetar's natal supernova dissipates a fraction $1 - W_{blowout} \sim 0.3 $ of the
initial energy injected, with voltages in the freely expanding wind available for particle
acceleration only after spindown from $\Omega_i$ to $\Omega_i W_{blowout}^{1/4} $, suggesting
a more realistic upper cutoff for the spectrum to be at
$E_{max}^{(eff)} = E_{max} W_{blowout}^{1/2} = 
8 \times 10^{21} \eta_1 \mu_{33} \Omega_4^2 (W_{blowout} / 0.7)^{1/2} $ eV.

Figure 1 illustrates another robust conclusion, that
the highest energy part of the spectrum is sensitive to
possible spindown losses due to gravity waves. The shortening of the
time in which particles can be accelerated to energies above $E_g$  suggests the
interesting possibility that observations of UHECR at energies
above $10^{20}$ eV can point the way
to sources of gravitational radiation, and vice versa (see \S \ref{sec:gravity_waves}).

While the agreement of this simple model with the observations is encouraging,
one must consider the possible modifications due to losses. Blowout from the source (in
this case, the supernova initially containing the relativistic wind),
radiative and adiabatic expansion losses are the most prominent pitfalls
for any compact object model for  high energy particles observed in any part
of the cosmic ray spectrum. Likewise, modification of the spectrum due to possible
diffusive transport in the intergalactic medium must be considered. I consider scattering 
in intergalactic magnetic fields first.

\subsection{Effect of Particle Scattering in an Integalactic Magnetic Field 
   \label{sec:scatter}}

There is no positive information pointing to the existence of a general intergalactic
magnetic field. Various limits suggest the magitude of such a field does not
exceed $\sim 10^{-9}$ Gauss ({\it e.g.}, Barrow {\it et al.} 1997), although 
a number of authors have argued
for larger fields injected from galaxies and spread around by ill-understood
processes ({\it e.g.}, Farrar and Piran 2000, Kronberg {\it et al.} 2001); 
see also Eilek and Owen (2002). 
If such a field exists, and if it causes high energy particles to
flow diffusively through intergalactic space (Adams {\it et al.} 1997), significant 
alterations of the UHE spectrum received from sources in galaxies would occur.

Consider what happens in a magnetic field strong enough, {\it and with sufficient magnetic
fluctuations,} to cause particles to have a scattering mean free path small compared
to the GZK length at energy $E$.  Assume the intergalactic field and its fluctuations
fill all space (no intermittency), The mean free path for a particle of energy $E$
in the fluctuating field is (Kulsrud and Pearce 1969)
$\lambda_B \approx r_L (E) (B /\delta B)^2, $
where $ r_L(E) =  E/ZeB = 341 E_{20.5} /ZB_9 $
Mpc, and
$ \delta B^2 \equiv  \langle k\delta B_k^2 \rangle_{kr_L = 1}.$
Here $E_{20.5} = E/10^{20.5} $ eV and $B_9 = B /10^{-9}$ Gauss.  If 
$\lambda_B (E) \ll l_{loss}$ (see expression \ref{eq:gzk_rate}), particles
come only from a distance 
$D_{scatter}(E) \approx \sqrt{\lambda_B l_{loss}} < l_{loss}$.
This diffusive reduction in the observable volume can occur at the highest energies,
including those above the saturation energy of the pion losses 
$E_{G} = 2.8 \times 10^{20}$ eV, only if
the general intergalactic magnetic field is strong, $B > 2 \times 10^{-8}$
Gauss, and the magnetic fluctuations are intense, $\delta B /B \sim 1 $, conditions
perhaps unlikely for a general intergalactic field.  

But at lower energies, the volume contributing to the observed particle flux
shrinks  when scattering is significant, compared to that observed if scattering
is negligible. This shrinkage depends on particle energy, which 
means that  diffusive transport can have a significant effect
on the observed particle spectrum between the ankle at $E_a = 6$ EeV and
$E_B = 1.1 \times 10^{20} (ZB_9)^{1/3} (\delta B/B)^{2/3}$ 
eV.  The shape of  the spectrum can be readily estimated in such a diffusive
transport regime. The
distance  $D(E) $ from sources to the observer is
$D_{eff} = (\lambda_B (E) cT_{loss}  )^{1/2} = cT_{eff} < cT_{loss} $. 
This contraction of the
observable volume leads to the spectrum in (\ref{eq:spectrum}) becoming
$J(E) \approx (c/4\pi) T_{eff} n_g \nu_m dN_i /d(m_i c^2 \gamma)$ $\propto E^{-3/2} $ 
instead of $\propto E^{-3} \Rightarrow E^3 J \propto E^{3/2}, $ 
in the energy range where
pion losses dominate, $E > 4 \times 10^{19}$ eV, instead of
$\propto E^0$,  contrary to the observations.
Solving the diffusion equation with systematic energy losses
in the short mean free path regime confirms this estimate.

Therefore, if there is a
general intergalactic magnetic field, this and all other metagalactic source models
which explain the absence of the traditional GZK cutoff by the existence of 
relatively nearby hard spectrum particle sources can work only if
$E_B < E_a$, which implies  $Z B_9 (\delta B /B )^2 < 1.5 \times 10^{-4} $ 
[equivalently, $\delta B / B < 0.01/(ZB_9)^{1/2} $] - either the 
general intergalactic field is weak, or the amplitude of the magnetic turbulence is
small, or both.

Compound diffusion, in which particles scatter strongly in localized regions,
with scatter free transport between the scattering 
sites, yields a similar contradiction with the data. If the average distance between 
scattering centers is an energy independent distance $l$ and the particles
lose negligible energy while trapped in the scattering sites, then 
$D_{eff} = (lc T_{loss})^{1/2} $ and and one finds $E^3 J \propto E $ between 
$E_a$ and the highest energy that can be trapped in a scattering site, also contrary
to observations of the well-determined spectrum at 10 to 50 EeV. 

The most likely
candidates for such localized scattering sites are galaxy clusters, with magnetic
fields perhaps  a few $\times 10^{-7}$ Gauss if smoothly distributed in the clusters
(see Eilek and Owen 2002 and references therein). The particles which
can be diffusively trapped in a (large) cluster have energy less than 
$10^{20} ZB_7 (\delta B /B)^2 $. Since $E^3 J $ is flat above $E_a$, one can conclude
that if UHE cosmic rays arise in normal galaxies with active star formation, as is
the case for the model proposed here, most of the particles get from the sources to us without
becoming diffusively trapped in large galaxy clusters - which is not surprising, since
galaxies with active star formation are not in rich clusters themselves, and rich clusters
occupy only a small fraction of the metagalaxy.  

\section{Acceleration and Escape of High Energy Particles from the Magnetar Wind and 
    the Magnetar Wind Nebula}

If magnetars were born naked, the theory would be complete, in terms of its energetics. 
The model's basic results appear in Figure 1 and 
Table \ref{tab:source_strength}.  For fiducial values of the input parameters
and weak or moderate gravitational wave losses, the theory does too
well - the source strength $K_0$ is too 
large, leading to the inference that only 5-10\% of the magnetars are born with voltages sufficient
to contribute to the UHECR.
The aim of this section is to estimate the losses imposed on the winds as they escape
their natal supernovae, to address the acceleration mechanism, and to argue that only half the
magnetars have magnetic geometry appropriate to their being  UHECR sources. The losses associated with
breakout from the supernova (estimated below to be $\sim$ 30\% of the initial rotation energy)
affect the energy of the upper cutoff of the spectrum; the geometric factors affect the
overall normalization. I also argue that in the equatorial outflow geometry appropriate to the model,
adiabatic and radiation losses are negligible.

\subsection{Escape of the Wind from the Supernova \label{sec:escape}}

Magnetars are neutron stars. Neutron stars form in core collapse
supernovae (Filippenko 2001 and references therein). The injection spectrum 
(\ref{eq:diff-inj-rate}) can form only if the
relativistic wind survives with negligible energy loss to large radii, {\it and}
only if the high energy particles accelerated in the wind can escape into interstellar
and intergalactic space with negligible adiabatic and radiation losses. In this and the 
following two sections, I outline the physics of the wind and particle
escape which justifies adoption of the bare magnetar 
spectrum (\ref{eq:spectrum}) as a reasonable ``bottoms-up'' model for the origin of UHE cosmic rays.

Particles with energy $E \sim E_{max}$, an energy
well above existing observations, must come from the 
outer limits of the relativstic wind, which at time $t$ lie between
$r = ct$ and $r = c(t-t_{spin,i})$, with $t_{spin,i} = \min(\tau_{EM}, \tau_{GW})$ - 
the outermost 5 light minutes or less.
If the energy in this region instead is dissipated through doing work on 
the plasma and radiation in the supernova and the presupernova environment, 
the spectrum would be truncated at the highest energies, and possibly suppressed
at lower energies if the environment dissipates wind energy at later times. 
The {\it observed} UHECR, with energies $E < 10^{20.5} E_{20.5}$ eV, require escape 
of the nonradiative wind to large radii only for
$\Omega <  3 \times 10^3 (E_{20.5} /Z\eta_1 \mu_{33} )^{1/2} \; {\rm s}^{-1}$; 
from (\ref{eq:simple_spin_hist}), such rotation rates occur 45 minutes and
more after the magnetar's formation.

The essential issue
arises in the initial spindown time being short or comparable to the time
for the explosive ejection of the supernova envelope\footnote{Many of the topics
discussed in this section have also been considered by Inoue {\it et al.} 
(2002) using a rapidly rotating pulsar with a $10^{13}$ Gauss magnetic 
field, in the context of the supranova model for gamma ray bursts,  assumptions
which yield rotational generation of electromagnetic energy on a time
scale long compared to the supernova explosion itself.}. Magnetars, if born rapidly rotating,
necessarily get rid of much of their rotational energy before the extended envelope
of an isolated Type II SNe can be fully ejected, which opens the possibility
of the energy of the relativistic wind being dissipated in the dense
matter around the newly formed magnetar, rather than being made available
to relativistic expansion and particle acceleration at large radii.

If core collapse SNe form in binaries, or shed their extended envelopes
before the core collapses through large mass loss, the presupernova
star may be the compact ($R_* \sim 10^{5.3}R_{5.3}$ km) 
stripped helium core of
the pre-supernova massive star, with envelope mass after
core collapse $M \sim 4 M_4$ solar masses
(Wheeler {\it et al.} 2000, Mazzali {\it et al.} 2002).
Compact progenitors
for core collapse SNe occur in Type Ib and Ic supernovae, and it is
interesting to note that the poorly known Ib/c supernova rates (Caparello {\it 
et al.} 1999) are comparable 
to the even more poorly known magnetar birth rate (Gaensler {\it et al.} 2001).
Then the dynamical time of the envelope surrounding the newly 
formed neutron star is fairly short, 
$t_d \sim 2.3 R_{5.3}^{3/2}/M_4^{1/2}$ minutes. The 
gravitational binding energy of the envelope is small,  
$W_g \sim 2 \times 10^{50} M_4^2 /R_{5.3}$ ergs, a figure which includes
the contribution from the neutron star's gravity. If there were no macroscopic
electromagnetic energy input from the newly formed magnetar, the supernova
shock, generated either from core bounce or neutrino driven convection
and expanding through the envelope at speed $v_s$, 
would begin ejection of the envelope in 
$t_{SN} = R_*/v_{s} \sim   18 R_{5.3} /v_{s30} $ 
seconds $ \ll t_d $, where
$v_{s30} = v_{s}/30,000$ 
km/s. The envelope thus forms a moving cavity containing the magnetar, expanding
with energy $\sim  4 \times 10^{52} M_4 v_{s30}^2$ ergs (Wheeler
{\it et al.} 2000).

The newly formed magnetar initially emits electromagnetic energy at a prodigious 
rate - $\dot{E}_{EM} \approx 2 \times 10^{50} \mu_{33}^2 \Omega_4^4 $ ergs/s, declining
to $\sim 10^{47}$ ergs/s after a few days.
Thus, within the time used by the supernova shock
to set the envelope in motion, the spinning magnet in the middle fills 
the cavity with  ``lightweight'' relativistic energy, mostly in 
the form of electromagnetic fields, since at this early phase radiation
losses bleed off the energy of any accelerated particles\footnote{The photon
energy created contributes to the lightweight energy inside the cavity, 
until the Rayleigh-Taylor instabilities which shred the envelope allow {\it all} 
the photons, electromagnetic fields and relativistic particles to escape into the 
surrounding circumstellar/interstellar (ISM/CSM) medium.}. Filling the cavity with
buoyant ``lightweight'' energy leads to prompt ``shredding'' of the envelope
through Rayleigh-Taylor instabilities.

Such shredding happens quickly.  Consider the ``radiation pressure'' exerted
on the envelope by the electromagnetic fields spun off from the newly formed magnetar.
This pressure creates an effective gravity at the expanding envelope's inner surface
$g \equiv U_{EM} A_{envelope} / M_{envelope}$. Since this acceleration of the
envelope's inner boundary is Rayleigh-Taylor unstable, buoyant ``bubbles'' of
light weight energy rise through the shell at the generalized 
Alfven speed $v_A = (U_{EM}/\rho_{envelope})^{1/2} \approx [\Delta E_{EM} (t)/M]^{1/2}$
(adapting the results of Davies and Taylor 1950 and Layzer 1954) and burst out, shredding the
envelope in the time $t_{shred}$.  
The total electromagnetic energy contained after spin down to angular velocity $\Omega $ is 
\begin{equation}
\Delta E_{EM} = \frac{1}{2} I\Omega_g^2 \ln 
    \left (\frac{1 + \frac{\Omega_i^2}{\Omega_g^2}}
          {1 + \frac{\Omega^2}{\Omega_g^2}} \right) \equiv 
          \frac{1}{2} I\Omega_g^2 \ln \Lambda(\Omega ) \approx
          1 \times 10^{52} \frac{\mu_{33}^2}{I_{45} \epsilon_2^2} \frac{\ln \Lambda}{2.4}
          \; {\rm ergs}.
\label{eq:EM_lost}
\end{equation}
For $\Omega \sim \Omega_g (\epsilon_2 =1), E \sim 10^{20.5} $ eV, $\ln \Lambda \approx 1.7$,
while at late times ($\Omega \ll \Omega_g, E \ll 10^{20.5}$ eV) 
$\ln \Lambda \approx \ln \Lambda (\Omega = 0) \approx 2.4$.
Expression (\ref{eq:simple_spin_hist}) gives an adequate approximation
for the time to spin down to angular velocity $\Omega$; the general expression is in
(\ref{eq:time_to_E}). Interstellar medium constraints 
(see \S \ref{sec:supershells}) suggest gravitational radiation
losses are substantial, thus requiring $\Omega_g < \Omega_i$ and the use of expression
(\ref{eq:EM_lost}) in evaluating the electromagnetic input to the supernova.

Rayleigh-Taylor instability, an adiabatic process, 
typically focuses the heavy fluid (the material in the envelope) into 
elongated ``spikes''  presenting a cross sectional
size $\sim R/\kappa $ with respect to the outflow,
covering a fraction $\sim 1/\kappa^2$  
of the initially unstable surface. The dominant scale of the
magnetic Rayleigh-Taylor instability (Kruskal and
Schwarschild 1954, Hester {\it et al.} 1996) as it becomes nonlinear is
the largest unstable mode that will fit in the thick envelope, 
$\lambda_i \sim R /2 $. Therefore, the number of envelope fragments is 
$N_{fragment} \sim A_{envelope}/\pi \lambda_i^2 \sim 16$.
Electromagnetic compression rapidly collapses the fragments into a small percentage of the volume 
- simulations of magnetic Rayleigh-Taylor instabilities
({\it e.g.}, Jun {\it et al.} 1995) suggest the fragments would have typical
dimensions  $\sim 0.5 \lambda_i \; (\kappa \sim 4)$ .

Envelope shredding
happens when the bubbles break through the expanding stellar envelope.
The bubbles rise faster in the expanding envelope as the electromagnetic energy accumulates.
Most of the envelope's material is in a shell of thickness $\Delta R $ not greater than the
presupernova star's radius $R_*$. 
Therefore  $t_{shred} \sim \Delta R / v_A (t_{shred}) \approx R_* /v_A (t_{shred})$.
Assume $\tau_g \gg t_{shred} \geq \tau_{GW}$, an assumption found to be 
self-consistent {\it a posteriori}. Then $\ln \Lambda \approx 1$ in (\ref{eq:EM_lost}),  
$\Delta E_{EM}(t_{shred}) \approx (1/2) I \Omega_g^2 \ln \Lambda 
 = 4.5 \times 10^{51} (\mu_{33}^2/I_{45} \epsilon_2^2 ) \ln \Lambda$,
$v_A \approx  7500 (\mu_{33}/\epsilon_2 )\sqrt{\ln \Lambda /I_{45}M_4}$ km/sec
and
\begin{equation}
t_{shred}  \approx 27 R_{5.3} \frac{ \epsilon_2}{ \mu_{33}}
   \left( \frac{M_4 I_{45}}{\ln \Lambda} \right)^{1/2} 
     \; {\rm seconds};
\label{eq:shred_time_GW}
\end{equation} 
this estimate assumes the envelope thickness is comparble to its radius.

The energy
$\Delta E_{shred} $ 
expended in shredding the shell into fragments 
which cover a fraction
of the area around the newly forming relativistic bubble
is not large.
The mechanical work that has to be done to shred the shell is
\begin{eqnarray}
\Delta E_{shred}  & \approx & M_{envelope} g v_A(t_{shred}) t_{shred} 
\approx \Delta E_{EM} \frac{v_A}{v_s} \nonumber \\ 
   &  = & \Delta E_{EM} \left(\Delta E_{EM} /M v_s^2\right)^{1/2}|_{t=t_{shred}}
     \equiv (1 - W_{shred} ) \Delta E_{EM}|_{t=\infty}.
\label{eq:E_shred}
\end{eqnarray}  
I used $g = U_{EM} A /M \approx \Delta E_{EM} /M R(t) \approx \Delta E_{EM} / M v_s t $ 
to obtain this expression.

Shredding of the envelope leaves most of the electromagnetic energy free 
to escape into the surrounding interstellar medium. From expression (\ref{eq:E_shred})
the fraction
\begin{equation}
W_{shred} = 1 -  \frac{\Delta E_{shred}}{\Delta E_{EM}|_{t=\infty}} 
     = 1 - 0.1 \frac{2.4}{\ln \Lambda (\Omega =0)}
               \left(\frac{\mu_{33}^2}{I_{45}M_4 \epsilon_2^2 v_{30}^2} \right)^{1/2}
\label{eq:W_shred}
\end{equation}
of the 
electromagnetic energy can escape to form a relativstic wind and Magnetar Wind Nebula, 
assuming mixing of the nonrelativistic material from the shredded supernova envelope
does not create too much mass loading.

The relativistic fields and particles emitted by the magnetar behave as a
relativstic ``fluid'' only so long as the non-relativistic material in the envelope
remains unmixed with the relativistic gas.
The Rayleigh-Taylor ``fingers'' are themselves subject to disruption from subsequent
Kelvin-Helmholtz instabiltities, whose final consequence is to mix the non-relativistic
material of the presupernova star's envelope with the relativistic outflow
{\it immediately in contact} with the nonrelativistic material, creating {\it local} ``mass
loading''. The Kelvin-Helmholtz mixing speed when one of the fluids is relativistic
is (Arons and Lea 1980) 
$v_{mix} \sim 0.1 \gamma_{KH} /k \approx 0.1  (U_{EM}/\rho_{finger})^{1/2} = 
.1 v_A /\kappa^{3/2}$, where $v_A = (\Delta E_{EM} /M)^{1/2}$ as in the discussion of 
interchange motions.  For the environment considered here, the mixing times
are on the order of an hour, during which time the envelope fragments move
to several hundred times the size of the presupernova star, while the 
relativistic gas not in contact with the fragments escapes at the speed of 
light. The nonrelativistic matter mixes
into only a small fraction, $\sim \kappa^{-2} \sim$
10-20\%, of the relativistic outflow
streamlines, thus leaving almost all of the relativistic energy that survives 
envelope shredding and 
fed in by the magnetar at later times to be free of
mass loading and able to accelerate high energy particles at radii large compared to
the size of the ejecta. The fraction of the energy surviving  mass loading thus is
given by the covering factor of the shredded ejecta, 
$
W_{load} \approx 1 - N_{fragment}(A_{fragment}/A_{envelope})
  = 1 -(N_{fragment}/4 \kappa^2) \sim 0.75.
$
Thus, $W_{blowout} = W_{shred}W_{load}  \sim  0.6-0.7 $; 
60-70\% of the lightweight
energy survives to break out of the confining supernova 
to blow a cavity in the surrounding interstellar/circumstellar 
medium (ISM/CSM),  before the envelope 
ejecta get a chance to drive the standard supervova blast wave.
Instead, the relativistic bubble (large scale,
low frequency EM fields and particles) drives a {\it relativistic} 
blast wave into the ISM/CSM. 

The losses involved in the relativstic wind
pushing out of the supernova come from the initial rotational energy
loss put into the wind.  Therefore, the highest voltage parts of the wind would be lost, which
would reduce the maximum particle energy that can be achieved by a factor
$W_{blowout}^{1/2} \sim 0.8 $.

\subsection{Relativistic Blast Wave and Magnetar Wind Nebula in the Interstellar
    Medium \label{sec:blast}}

The particles
accelerated in the wind still must escape the Magnetar Wind Nebula without 
appreciable loss, even after the wind and the MWN escape the supernova, if the
bare magnetar spectrum  (\ref{eq:diff-inj-rate}}) is to represent the rate ultra-high
energy ions are injected into the metagalaxy.
For times greater than $t_{shred}$, the relativistic energy escaping from the
supernova drives a shock wave directly into the surrounding interstellar
or circumstellar medium, superseding the usual shock driven by the supernova ejecta -
the ejecta are left behind.  For simplicity, I consider
only the case of an explosion into a uniform external medium of mass density
$\rho_1$; the results for
an explosion into a cavity excavated by mass loss from the supernova's progenitor,
with mass density $\propto r^{-a}$, are similar to the uniform case. 

Since $\Delta E_{EM} / (ct)^3 \gg \rho_1 c^2 $ for many months after the explosion,
for all reasonable densities of the interstellar medium, the relativistic energy
drives a relativistic blast wave, whose Lorentz factor  and radius are
\begin{eqnarray}
\Gamma_b & = &\left[\frac{17}{8\pi} \frac{\Delta E_{EM}}{\rho_1 c^2 (ct)^3} \right]^{1/2}
    \label{eq:blast_gamma} \\
R_b & = & c\left[t - \int_0^t dt'\frac{1}{2\Gamma_b^2 (t')} \right] 
   = ct\left[ 1 - \frac{1}{8\Gamma_b^2 (t)} \right].
    \label{eq:blast_radius}
\end{eqnarray} 
The dependence of $\Gamma_b$ on $\Delta E_{EM}$ and $\rho_1$ comes from 
simple consideration of momentum conservation; the factor $17/8\pi$ comes
from the similarity solution (Eltgroth 1972, Blandford and McKee 1976). The
shocked interstellar medium occupies a thin shell, whose thickness is
$\Delta R_{ism} = R_b/6\Gamma_b (t)$.  Within this thin shell, the interstellar
magnetic field is greatly amplified, 
$
B_{2ism} = 3 B_1 \Gamma (t).
$
Flux conservation since the start of the blast yields the coeffcient $3$ in
this expression; the instantaneous jump conditions replace $3$ with $2$, which
applies to the $B$ field just behind the blast wave. The relativstic expansion continues
until the time $T_{rel}$ such that $\Gamma_b (T_{rel}) \sim 1$, which yields
\begin{equation}
T_{rel} = \left( \frac{17}{8\pi} \frac{\Delta E_{EM}}{\rho_1 c^5} \right)^{1/3} =
   1.75 \left(\frac{\Delta E_{52}}{n_1} \right)^{1/3} \; {\rm years};
\label{eq:Trel}
\end{equation}
$n_1$ is the baryon number density upstream of the blast wave in units of 1 cm$^{-3}$.
For $t > T_{rel}$, the blast enters the nonrelativistically expanding, energy
conserving Sedov phase.

The relativistic fields and particles from the magnetar fill the volume out
to a contact surface located at $R_c = R_b - \Delta R_{ism}$, which is close
to $R_b$ when $t < T_{rel}$; it suffices to take $R_c$ as being equal to
$R_b$.  This volume forms the Magnetar Wind Nebula (MWN)
in its relativstic expansion phase. The freely expanding magnetar's wind terminates in
a relativistic reverse shock wave located at radius $R_s$ where the momentum flux in the wind 
matches the pressure in the magnetar wind nebula (almost all of which is in the 
initial energy pulse $W_{blowout}\Delta E_{EM}$). I consider only times greater
than $t_{shred} > \tau_{GW}$, since at earlier times the relativistic blast wave has not
yet formed.

The pressure balance condition is
\begin{equation}
\frac{\dot{E}_{EM} (t - R_s /c )}{4 \pi R_s^2 c} = U_{MWN} \approx 
      \frac{3}{4\pi}\frac{\frac{1}{2}I \Omega_g^2 W_{blowout} \ln \Lambda (\Omega)}{R_b^3}.
\label{eq:press_bal}
\end{equation}
One readily finds, upon using (\ref{eq:simple_spin_hist}) and assuming $R_s$ is not almost
equal to $ct$, that the reverse shock forms where
\begin{equation}
\left( \frac{R_s}{R_b} \right)^2  =  
     \frac{2}{3 W_{blowout} \ln \Lambda} \left(\frac{2 \tau_g}{t} \right)^{4p-1}
\label{eq:Rs-Rb}
\end{equation}
with $p = 1/4$ for $\tau_{GW} \ll t < 2 \tau_g$, and $ p = 1/2 $ for $t > 2 \tau_g$. At early
times, when gravity waves control the spindown, the reverse shock expands 
with the MWN ($4 p - 1 =0, \; R_s = (2\sqrt{W_{blowout} \ln \Lambda}/3) ct
\approx 0.63 ct$), 
but for times greater than $2 \tau_g, \; p =1/2$ and the blast wave leaves the
reverse shock behind\footnote{If gravity wave losses are negligible, pure electromagnetic spindown
yields
$(R_s/R_b)^2 = (2/3 W_{blowout}) (2\tau_{EM}/t), \; 
R_s = \sqrt{2  /3 W_{blowout}} c(2 \tau_{EM} t)^{1/2} = 4 \times 10^{13} 
    (I_{45}/\mu_{33}^2 \Omega_4^2 )^{1/2} t_{hours}^{1/2} \; {\rm cm}$
when $t \gg 2 \tau_{EM} \gg t_{shred}$.  Since $t_{shred} \ll 2 \tau_{EM}$, at very early times  
$R_s \approx (2/3 W_{blowout})^{1/2} ct = 3 \times 10^{13} (0.7/W_{blowout})^{1/2} 
 (t/ 10 \: {\rm min} ) \; {\rm cm}, \;t < 2 \tau_{EM} $.}, 
with
\begin{equation}
R_s = \left(\frac{2}{3 W_{blowout} \ln \Lambda }\right)^{1/2} c(2t\tau_g)^{1/2}
     =  6.5 \times 10^{13} \left(\frac{0.7}{W_{blowout}} \frac{2.4}{\ln \Lambda}
         \frac{I_{45}^3 \epsilon_2^2}{\mu_{33}^4} \right)^{1/2} t_{hours}^{1/2} \; {\rm cm}.
\label{eq:Rs}
\end{equation}

Because the wind, at $r < R_s $, and the MWN, between the $R_s$ and $R_c$,  
both expand relativistically, the electromagnetic fields 
{\it everywhere} ($0 < r < R_c $) have amplitude proportional to $1/r$, 
\begin{equation}
B (r,t) = \sqrt{\frac{\sigma}{1+\sigma}} 
                 \frac{\mu \Omega^2 (t-r/c)}{c^2 r} f(t-r/c)
        = \sqrt{\frac{\sigma}{1+\sigma}}  \frac{\Phi(t-r/c)}{r} f(t-r/c),
\label{eq:Bwind}
\end{equation}
where $\sigma$ is the ratio of the electromagnetic energy flux to the plasma
kinetic energy flux in the wind interior to $R_s$. 
$f(r,t)$ describes short wavelength [$\lambda \approx 2\pi c/\Omega (t - r/c)$] structure in the
wind, as is the case if the wind has wave structure similar to vacuum
strong waves or, if an MHD wind, is ``striped'' (Michel 1971, Coroniti 1990, Michel
1994, Bogovalov 1999), with the B field direction reversing with wavelength $\lambda$ 
while the overall amplitude declines inversely with $r$.

Pulsar Wind Nebulae (PWN) observed in our own galaxy expand nonrelativistically, 
due to the fact that the initially injected relativistic energy $\Delta E_{EM}$
is small compared to the kinetic energy of the conventional supernova. The resulting
``bags'' of
electromagnetic energy and particles injected by the pulsar are confined by the 
inertia of the supernova ejecta until times large compared to their ages 
when we see them, as is the case in the Crab Nebula (Hester {\it
et al.} 1996), or by the surrounding CSM or ISM, as might be the case for the
nebula around PSR 1509-58 ({\it e.g.}, Gaensler {\it et al.} 2002). In contrast,
the hypothesized MWN described here expands relativistically at all radii 
$R_s \leq r \leq R_c = R_b - \Delta R_{ism} $, for $t < T_{rel}$.  Therefore,
the post-shock velocity at $R_s$ must be relativistic, and $\sigma \gg 1$ - the 
magnetar's wind and the MWN must be everywhere electromagnetically dominated and
``clean'' - throughout the relativstic expansion phase, the field structure will
not be disturbed by interaction with clumps of nonrelativstic matter, since all
such material lies in the shreds of supernova material through which
the wind escapes, and  in the shocked shell of ISM/CSM right behind the blast wave.

If magnetars behave like pulsars and have wind outflows with a dense $e^\pm$ plasma
$(\dot{N}_\pm \gg \dot{N}_{GJ}$) in addition to the Goldreich-Julian fluxes 
associated with the electric currents
(the high energy ions form one of these electric current flows), the wind and MWN
would have a structure perhaps described by MHD.  $\dot{N}_\pm$ for a magnetar
is not well understood, either theoretically 
or observationally. Observed magnetars
are slow rotators with voltages below the level at which pulsars stimulate
observable PWN, so prominent nonrelativistic MWN around observed galactic magnetars would not
be expected. Theoretically, pair production in the ultrastrong magnetic fields
of magnetars has been controversial, with the inhibitions of pair creation by
photon splitting being uncertain (see Baring and Harding 2001). However, since 
$\sigma $ must be large, the MWN would not be an efficient radiator, if all the
acceleration of pairs is associated with the reverse shock at $R_s$, until 
$t > T_{rel} \sim $ 1 year.

\subsection{Acceleration of UHE Ions in the Magnetar Wind \label{sec:accel}}

The metagalactic magnetar model for UHE cosmic rays has been developed using 
the phenomenology and theory of relativistic winds in galactic Pulsar Wind Nebulae
(PWN) as the basic physics input. So far I have not described a specific acceleration mechanism, 
through which the ions in the equatorial Goldreich-Julian current actually tap the voltage
available in the wind; the same lacuna exists in the theory of ions in PWN.  A detailed
investigation of this subject is outside the scope of this paper. However, a few remarks
are relevant, addressed to the most likely mechanism in this context, the ``surf-riding''
of charged particles in the relativistically outflowing electromagnetic fields.

As is clear from expression (\ref{eq:wind_volts}), the ions would gain energy comparable to the
total available voltage if they can cross magnetic field lines for a distance comparable to one 
decade in radius.
Such field line crossing can occur in either radius or in latitude, or both. 
To see how the simplest form of surf-riding works,
consider particle motion in the asymptotic wind of the aligned rotator (Buckley 1977,
Contopoulos and Kazanas 2002).

Substantial theoretical evidence has accumulated that when the electromagnetic fields dominate the
outflow's energy density, the structure of the dipolar magnetosphere and its outflow can be 
approximated as force-free [$\rho_q {\bf E} + (1/c){\bf J \times B} = {\bf 0}$, 
with $\rho_q$ the charge density and ${\bf J}$ the current density], and that under the 
magnetohydrodynamic conditions
appropriate to PWN ($n \gg \rho_q/q$, where n is the total quasi-neutral plasma density),
the structure of the aligned rotator's 
wind at radii larger than a few $R_L$
closely approximates that of the split monopole (Michel 1974 , 
Contopoulos, Kazanas and Fendt 1999), except perhaps in narrow cones around the 
rotation axis. These fields have 
the form (Michel 1973, Bogovalov 1997)
\begin{eqnarray}
B_r & = & \pm \frac{M}{r^2}, \;  B_\phi  =  \mp \frac{M \sin \theta}{rR_L} \nonumber \\
E_\theta & = & B_\phi \label{eq:fields} \\
B_\theta & = & E_r = E_\phi = 0. \nonumber
\end{eqnarray}
The magnitude of the monopole moment is $M = k \mu /R_L$, where $\mu $ is the dipole moment
and $k$ is a constant on the order of unity.  For the self-consistent numerical solution
obtained by Contopoulos {\it et al.} (1999), $k = 1.36$.
The coordinate system is spherical, with $\theta = 0$ being the colatitude of the angular momentum
vector, and with the azimuth $\phi$ being measured with respect to an arbitrary $x$ axis.
The plus and minus signs apply to the opposite (northern/southern or southern/northern, 
with respect to the rotation axis) hemispheres. The fields reverse direction abruptly across the
rotational equator ($\theta = \pi /2$), which implies the existence of an equatorial current sheet 
where the return current flows. In the body of the wind, the volume current is radial.

The electromagnetic energy propagates with the field line velocity 
\begin{eqnarray}
{\boldsymbol v}_E & = & c \frac{{\boldsymbol E \times B}}{B^2}
  = c \left(\hat{\boldsymbol{r}} \frac{x^2}{1 + x^2} + \hat{\boldsymbol{\phi}} \frac{x}{1+x^2} \right)
     \label{eq:v_E} \\
v_E & = &|{\boldsymbol v}_E | = c \frac{x}{\sqrt{1 + x^2}}
    \nonumber \\
\gamma_E & = & \left[ 1 - \left(\frac{v_E}{c} \right)^2 \right]^{-1/2} = ( 1+x^2)^{1/2}, \; 
      x \equiv \frac{r\sin \theta}{R_L}  \label{eq:gamma_E}.
\end{eqnarray}
Charged particles flow out with with velocity 
$\boldsymbol{v} = \boldsymbol{v}_E + v_\parallel \boldsymbol{b}$, where 
$\boldsymbol{b}$ is the unit vector along the magnetic field. When particles enter
the wind with initial energy $m c^2 \gamma_L$ (corresponding to initial flow 4-velocity 
$c \beta_L \gamma_L, \; \gamma_L = (1-\beta_L^2)^{-1/2}$), their parallel speed is,
when $\gamma_L \gg 1 $ (Arons, in preparation) 
\begin{equation}
\beta_\parallel = \frac{\gamma_L}{\gamma_E \sqrt{\gamma_L^2 + \gamma_E^2}}.
\end{equation}
Therefore, the wind's Lorentz factor is
\begin{equation}
\gamma_w = (1 - \beta_E^2 - \beta_\parallel^2)^{-1/2} = (\gamma_L^2 + \gamma_E^2)^{1/2}
    = \left[1 + \gamma_L^2 +\left(\frac{r}{R_L} \right)^2 \sin^2 \theta \right]^{1/2}
\label{eq:large_energy}
\end{equation}  
(Contopoulos and Kazanas 2002).  When $r \sin \theta \gg \gamma_L R_L$,
$\gamma \rightarrow  r \sin\theta /R_L$ (Buckley 1977); the wind is a linear
accelerator, with each particle's energy increasing by one factor of 10 for each
decade in radius. At large radii, the particles move with the field lines, with
negligible motion parallel to $\boldsymbol{B}$, that is, they ``surf-ride'' on the
electromagnetic field.

This result applies to quasi-neutral winds, whose total density
is large compared to the Goldreich-Julian density; it does not apply to the charge separated
force free split monopole (Arons, in preparation). Pair creation within the magnetosphere
supplies densities large compared to $n_{GJ}$, with injection energies $\gamma_L \sim 10 - 10^3$;
the precise values of $n$ and $\gamma_L$ depend
on the specifics of the magnetospheric parameters (surface magnetic field strength, voltage
and whether curvature or inverse Compton gamma rays radiated by electrons or positrons
accelerated within the magnetosphere initiate the pair cascades.)

Such surf-riding acceleration may seem surprising, since ${\boldsymbol v \cdot E} = 0$.  
The physics of the acceleration appears when one considers the (small) effect
of inertia on the velocity transverse to $B$, by solving the equation of motion to first
order in the rest mass $m$. The solution for the velocity perpendicular to $\boldsymbol{B}$ 
at large radius, where the velocity becomes equal to the radial component of the
$\boldsymbol{E \times B}$ drift, is
\begin{equation}
{\boldsymbol v} = {\boldsymbol v}_{Er} + \frac{mc}{q} \frac{ {\boldsymbol B} 
     \times D(\gamma {\boldsymbol v})/Dt}{B^2}
\approx {\boldsymbol v}_{Er} + \frac{mc}{q} \frac{{\boldsymbol B} 
     \times D(\gamma_E {\boldsymbol v}_{Er})/Dt}{B^2}
       = {\boldsymbol v}_{Er} + {\boldsymbol v}_{pol}.
\end{equation}
Here $D/Dt = \partial /\partial t + {\boldsymbol v \cdot \nabla} \approx 
\partial /\partial t + v_{Er} ( \partial /\partial r)$.
The polarization drift velocity ${\boldsymbol v}_{pol}$ is parallel to the electric field:
\begin{equation}
{\boldsymbol v}_{pol} = \frac{mc}{qB^2} {\boldsymbol B} \times 
      \frac{D(\gamma_w {\boldsymbol v}_{Er})}{Dt}
   = \mp \frac{mc^3}{q }\frac{r}{M} \hat{\boldsymbol{\theta}} ,
\label{eq:v_pol}
\end{equation}
where $\hat{{\boldsymbol \theta}} $ is the unit vector in the $\theta$ direction, and I have assumed
$r \sin \theta \gg R_L $. ${\boldsymbol v}_{pol}$ always has a direction such that 
${q\boldsymbol v}_{pol}\cdot {\bf E} $ is positive - particles gain energy from the electric field
in an accelerating flow.
Then energy conservation takes the form
\begin{equation}
mc^2 \frac{D \gamma}{Dt} = q {\boldsymbol v}_{pol} \cdot \boldsymbol{E}_\theta 
   = + \frac{mc^3  \sin \theta} {R_L}.
\end{equation}
Integrating with $D/Dt = c ( \partial /\partial r)$ yields expression (\ref{eq:large_energy})
in the limit $r \sin \theta \gg  \gamma_L R_L$.  This correspondence can be extended to all
radii, at the expense of much more complex algebra.

Thus relativistic force free winds can be linear accelerators, with the particles ``surf-riding''
on the electromagnetic  fields.  For the aligned rotator, these fields form a (spherical)
DC transmission line, with the fluid accelerating following the increase of the energy transmission speed 
toward $c$ as the magnetic field approaches a purely transverse structure with increasing radius -
the wind 4-velocity $\gamma_w $ is not constant but is  asymptotially $ \propto r$.
For $c /\Omega = 30/\Omega_4 $ km and $\gamma_L \leq 10^3$, such linear acceleration sets in
at radii not larger than $(30,000/\Omega_4) (\gamma_L /10^3) $ km.

The result (\ref{eq:large_energy}) applies to force free conditions, which require
$\sigma \equiv B^2/4\pi \rho \gamma_w c^2 \gg 1$,
where $\rho$ is the rest mass density measured in the frame where the neutron star's 
center of mass is at rest.
Assume plasma at nonzero latitude $|\lambda | = |\theta - \pi/2 | > 0$ crosses the light cylinder
with initial energy/particle $mc^2 \gamma_L = E_L$ and rest mass density 
$\rho_L = (m_i + 2 \kappa_\pm m_\pm) n_{GJ}(R_L) \equiv m_{eff} \Omega B(R_L) /2\pi c q \approx 
m_{eff} \Omega \mu /2\pi c q R_L^3$, where I have assumed the heavy ions have rest mass
$m_i$ and density $n_i = n_{GJ}$ and are mixed with a pair plasma of density
$n_+ + n_- = 2 \kappa_\pm n_{GJ} \gg n_i$. Then 
\begin{eqnarray}
\sigma_L & = & \sigma(R_L) = \frac{B_L^2}{4 \pi m_{eff} (\Omega B_L /2\pi cq) \gamma_L c^2} =
   \frac{1}{2} \frac{qB_L}{m_{eff} c \gamma_L \Omega} = 
    \frac{1}{2} \frac{q \Phi_{mag}}{m_{eff} c^2 \gamma_L } \label{eq:sigma_L}, \\
m_{eff} & \equiv & m_i + 2 \kappa_\pm m_\pm.  \label{eq:m_eff}
\end{eqnarray}

In the absence of magnetic disspation and for $r \sin \theta \gg \gamma_L c/\Omega$, 
$\sigma = \sigma_L \gamma_L (R_L /  r \sin \theta)$, with the force free region ending
at $r =R_1 =R_L \gamma_L \sigma_L /\sin \theta $, where $\sigma$ drops to unity.  Beyond $R_1$, the
plasma probably coasts, with $\gamma_w = \gamma_1 \approx \gamma_L \sigma_L $, in regions where 
$\sin \theta \sim 1$. If the ions are absent, $m_{eff} \rightarrow  2 m_\pm \kappa_\pm$.

The polarization drift velocity (\ref{eq:v_pol}) causes the
ions (and other particles) to drift across B (that drift is the origin of the acceleration), 
moving off their radial orbits toward the rotation pole a distance 
$\Delta l (r) = (m_{rest} c^2 /q\Phi_{mag}) (r^2/R_L) $.  Since the acceleration
ends at radii no larger than $r = R_1$, the maximum fractional energy gain for an ion 
from polarization drift alone is
$E /q\Phi_{mag} =  m_i / m_{eff}$. If $m_{eff} \approx 2-10 m_i$ and $\gamma_L \sim$
a few (as is the case when pair creation cascades are strong, as in the young pulsars)
in the return current region, 10\% - 50\% of the magnetsopheric
potential on open field lines could go into accelerating the ions due to polarization
drift along $E$.  This fraction is lower if the pressure of the MWN
terminates the wind interior to $R_1$.

The accelerator closely approximates a purely radial motion - curvature 
(a.k.a. synchrotron) radiation losses are negligible, so long as the particles
surf-ride (${\boldsymbol v} \approx {\boldsymbol v}_{Er}$).

The energy that can be achieved within $r = R_1$ thus depends on the plasma composition.
For an aligned rotator with magnetic moment parallel to the angular momentum, the ions
in the wind would flow in the equatorial current sheet, while at higher latitudes, the
plasma would be composed solely of pairs. The theory of pair creation in ultra-magnetized 
neutron stars is not as well understood as in standard pulsars ($\mu \sim 10^{30}$ cgs), 
since a variety of strong field QED effects can intervene to inhibit pair creation. 
A traditional estimate of $\kappa_\pm $ which ignores these effects ({\it e.g.}, Ruderman
and Sutherland 1975, Arons and Scharlemann 1979) and that takes account of the limitations
set by curvature gamma ray emission on the energies of electrons in the primary polar particle
beam yields $\kappa_{\pm} \sim 10^4 (\mu_{33} R_{10}^2 )^{1/4}$. Taking account of photon splitting,
and pair production directly into the lowest Landau levels, effects that reduce the
contributions of the synchrotron cascade to pair creation (Baring and Harding 2001), suggests
$\kappa_\pm \sim 10-100$.
This polar flow would reach the light cylinder with Lorentz factor $\sim 10-100$, yielding
$E_1 = m_\pm c^2 \gamma_1 = m_\pm c^2 \gamma_L \sigma_L 
 = 3 \times 10^{21} \mu_{33} \Omega_4^2 (10/\kappa_\pm $
for the asymptotic 4 velocity of an ideal MHD, {\it pair dominated} wind 
outside the equatorial current sheet,
assuming no magnetic dissipation interior to $R_1$.

In the aligned rotator models contructed to date, the dynamics of charges in the current sheet,
where the ions flow as the return current, are not specified. Simple models of the
current sheet's structure (Arons, in preparation) suggest it to be a force-free
rotational discontinuity separating the hemispheres.  This structure causes
the ions to be frozen to the fields which partcipate in the general linear acceleration,
and therefore also obey the linear acceleration law (\ref{eq:large_energy}). If the geometry
has the dipole moment antiparallel to the rotation axis, then the ions are themselves the charge 
carriers in the $|\lambda | > 0 $ hemispheres and participate in the generaL acceleration, 
while the current sheet carries a negatively charged outflowing return current.

\begin{figure}[H]
\begin{center}
\begin{picture}(450,200)(0,15)
\put(0,0){\makebox(450,200){\epsfxsize=450\unitlength \epsfysize=200\unitlength
\epsffile{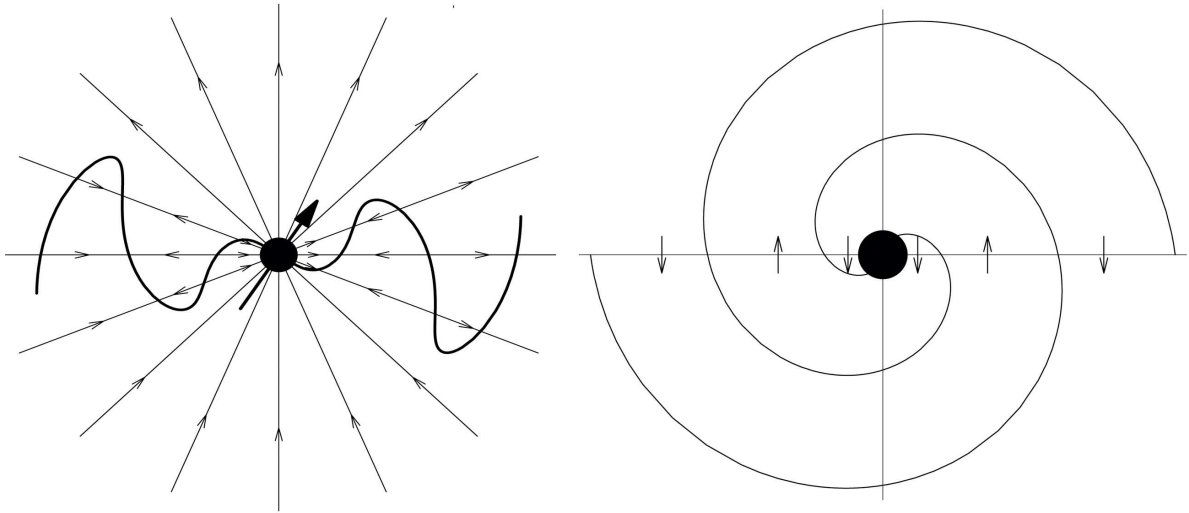}
}}
\end{picture}
\end{center}
\hspace{1\unitlength}
\vspace{-1\unitlength}
\caption{Frozen-in current sheet structure of the oblique split monopole in the 
inner wind, from Bogovalov (1999).
Left panel: Meridional cross-section of the poloidal field structure at large $r$,
showing the crinkled current sheet.   
Right panel: Intersection of the curent sheet with the equatorial plane.
The toroidal magnetic field forms stripes with opposite directions between each current
layer, as indicated by the arrows between the current sheets.}
\label{fig:split_mon}
\end{figure}

In general, magnetized rotators are not expected to have their dipole moments exactly aligned
or counter-aligned to the angular velocity. While a solution of the force-free oblique rotator
with dipolar field structure inside the light cylinder has not yet appeared, the
experience with the aligned rotator suggests the wind emerging from an oblique rotator 
will have the structure of the {\it oblique} split monopole at radii larger
than a few times $R_L$. An exact
solution for the oblique split monopole has been obtained by Bogovalov (1999), 
whose results for the fields are shown in Figure \ref{fig:split_mon}.

The magnetic and electric field strengths in the oblique split monopole are the same
as in the aligned split monopole; the only difference is that in the equatorial
region (latitudes $-i < \lambda < i$, where $i$ is the angle between the magnetic and
rotation axes), the directions of $E$ and $B$ reverse every half-wavelength, where the wavelength
is $2\pi R_L$, forming a ``striped'' wind (Michel 1971, Coroniti 1990).  The current sheet is
now wrinkled and is frozen into the flow, advecting outwards with the same 4-velocity as the
general wind. That 4-velocity is exactly the same as in the aligned split monopole,
given by (\ref{eq:large_energy}). Thus, under the plausible assumption that the oblique
dipole's wind takes the asymptotic form of the oblique split monopole, the 
maximum ion energy/particle achieved in all magnetic geometries of rapidly rotating magnetars 
through the action of polarization drift of the ions along the MHD electric field is
\begin{equation}
E_{max} = m_i c^2 \gamma_1 = m_i c^2 \gamma_L \sigma_L = q \Phi_{mag} \frac{m_i}{m_{eff}},
\label{eq:E_max}
\end{equation}
where $m_{eff} = m_i + 2 \kappa_\pm m_\pm$ and $\kappa_\pm$ is the pair multiplicity {\it in the
current sheet}.

If $\kappa_\pm \leq 10^3$ in the current sheet, $m \approx m_i $ and this elementary form of
surf-riding yields 
$\sigma_L = 0.5 Ze\Phi_{mag}/A m_p c^2 \gamma_L = 
6 \times 10^{12} (Z/A) \mu_{33} \Omega_4^2 (10/\gamma_L)$. Then 
$R_1 = R_L \gamma_L \sigma_L = 3.5 (Z/A) \mu_{33} \Omega_4^2 $ pc, and
$E_{max} = 3.3 \times 10^{22} Z \mu_{33} \Omega_4^2 $ eV. 

Linear acceleration due to polarization drift cannot not go to completion in the freely 
expanding ($r < R_s$) magnetar wind 
since $ R_s \ll R_1$. 
The MWN exterior to the large $\sigma$ shock ending the wind also expands relativistically
but with a {\it decelerating} field line velocity. Therefore, if the magnetar wind model
is to explain UHECR, there must be a still stronger acceleration
gradient in the wind; laminar surf-riding in the decelerating MWN saps particles 
of their energy\footnote{In pulsar wind 
nebulae, large-scale surf-riding in the wind {\it is} an adequate explanation of the 
infered wind four velocities (Contopoulos and Kazanas 2002, Spitkovsky and Arons 2003)}.

The polarization drift based linear accelerator is only a lower limit to the rate at which 
surf-riding in the magnetar's wind can can accelerate the charges.  If the fields
become dissipative within the wind, more rapid acceleration can occur. The low magnetization
inferred for the equatorial sectors of pulsars' winds suggests that such dissipation must
exist.  

Figure \ref{fig:split_mon}
shows that for an obique rotator, much of the Poynting flux is tied up in waves.  In the
ideal MHD outflow, these waves are frozen in current sheets. These structures become 
thinner as the wind expands, and are subject to
inductive dissipation (Coroniti 1990, Michel 1994, Lyubarsky and Kirk 2001, Lyutikov 2002) 
at and beyond the disipation radius of the crinkled current sheet
$R_{diss} \sim 2\kappa_\pm R_L \ll R_s$. They are also subject to instability
with respect to emission of {\it kinetic} Alfven waves (Bellan 1999, 2001, and 
Arons, in preparation) which propagate
with respect to the plasma in the wind frame. Current driven instability with respect to 
wave emission is a faster process than the inductive dissipation mechanisms discussed to date.

Assume wave emission is the dominant process dissipating the wrinkles in the current sheet, and that
such emission leads to the formation of large amplitude electromagnetic waves with 
dimensionless amplitude $a = q \delta B /m_i c \Omega \gg 1$. 
These waves can accelerate ions faster than does polarization drift in 
$E_\theta $. Even when $r < R_{diss}$, the Alfven waves can act as
powerful ponderomotive accelertors (Chen {\it et al.} 2002), while for $r \geq R_{diss}$ 
these waves become superluminal ``strong
waves'' (Melatos 1998), whose ponderomotive force also yields surf-riding acceleration.   When
the waves are kinetic Alfven modes, the ponderomotive force applies preferentially to the ions.

An estimate of the ponderomotive work done suffices for the present discussion.

This single particle force arises since the transverse motion of a charged particle in 
the wave electric
field gives rise to ${\boldsymbol v \times B} $ force in the wave magnetic field 
which points radially outwards, does not
average to zero over a wave oscillation, and has magnitude 
\begin{equation}
{\bf f}_{pond} = \frac{q}{c} \langle \delta {\bf v} \times \delta {\bf B} \rangle = 
   \frac{q}{mc} \left\langle \frac{\delta {\boldsymbol p} _\perp}{\gamma_{osc}} 
            \times \delta {\boldsymbol B} \right\rangle 
      =\frac{q^2}{mc\Omega} \left\langle \frac{\delta {\boldsymbol E}}{\gamma_{osc}} 
           \times \delta {\boldsymbol B} \right\rangle
               = mc\Omega \frac{a^2}{\sqrt{1+a^2}} \hat{{\bf r}},
\end{equation}
where a particle's transverse oscillation momentum in the strong wave is 
$\delta {\boldsymbol p}_\perp \approx q\delta {\boldsymbol E}/\Omega $, 
the oscillation energy per particle is 
$m_i c^2 \gamma_{osc} = m_i c^2 [1 + \langle (\delta p_\perp /m c)^2 \rangle ]^{1/2} = 
 m_i c^2 \sqrt{1 + a^2} $, 
$ \hat{{\boldsymbol r}} $ is the unit vector in the 
radial direction and $ \langle \; \rangle$ indicates an  average over wave phase. The work 
done on a particle as it moves a distance $\Delta r$
through the wave is then, when $a \gg 1$,
\begin{equation}
W_{pond} = f_{pond} \Delta r = (mc\Omega \: a) \Delta r
    = qBr \frac{\delta B}{B} \frac{\Delta r}{r} 
         \approx q \Phi_{mag} \eta, \; \eta \equiv \frac{\Delta r}{r},
\end{equation}
since $\delta E \approx \delta B $ and $\delta B /B \geq 1$ in the rotational equator - 
most of the energy flow is in the wave 
Poynting flux. The evidence from Pulsar
Wind Nebulae is that $ \eta \approx 0.1$ and therefore $\Delta r \sim 0.1 r$. 
The fact that $\Delta r \ll r$ and the particles gain only a relatively small part of the
total voltage available most likely is due to the demise of the strong wavesÕ coherence, as a result of 
parametric instabilities in these electromagnetic structures - see, for example, 
Leboeuf {\it et al.}. (1982). 

This argument suggests that such surf riding may be the source of the acceleration hypotheiszed
in the model for UHECR studied here. A detailed demonstration that this is so is a task for 
another study.  If so, the ion
acceleration is prompt, beginning a few seconds and continuing for a few days after the 
escape of the relativistic wind from the natal supernova - that escape occurs $\sim$ a minute after
the neutron star's formation - see expression (\ref{eq:shred_time_GW}).

The standard alternate for the acceleration mechanism of UHECR is some form of shock acceleration,
usually invoked as diffusive Fermi acceleration in relativistic shocks behaving in a
manner similar to nonrelativistic shocks ({\it e.g.},
Waxman 1995, Bahcall and Waxman 2003); see Ostrowski and Bednarz (2002) and references 
therein for a discussion and critique of this mechanism's plausibility in relativistic 
shocks.)  In the current context, the fact that the 
electromagnetic fields dominate the energetics of the outflow militates against 
the relevance of 
that mechanism. Alternate, ``flare'' mechanisms, based on direct dissipation of the
currents, are even more poorly understood.  
In this regard, identifying bursts of UHE cosmic rays as coming from the same direction as
bursts of quasi-coherent gravitational radiation, with correlated arrival times, would
give substantial empirical input into constraining the specific physical processes behind
the acceleration of the highest energy cosmic rays.

\subsection{Escape of UHE ions from the Magnetar Wind Nebula \label{sec:MWNescape}}

In accord with the discussion in \S \ref{sec:accel}, I assume the high energy ions achieve their energy
in the relativistic wind through surf-riding acceleration,
at a radius $R_a \sim R_{diss} \approx \kappa_\pm R_L = (10^3-10^4) R_L = (30,000-300,000)/\Omega_4$ km.
Beyond $R_a $, dissipation of the strong waves reduces the structure of the magnetic field to 
that of the aligned rotator (Blandford 2002), with fields given by (\ref{eq:fields}) but now with
$M \equiv \mu/R_L \rightarrow M_\parallel \equiv \mu \cos i /R_L$. The acceleration injects
the ions into the angular sector $-i < \lambda < i$, with a 4-velocity large compared to the
wind's 4-velocity (\ref{eq:large_energy}). One can readily show, either from the equation of motion
of a UHE ion in the crossed $E$ and $B$ fields of the wind, or from the motion of an ion
in the pure B field in the $E = 0$ frame of the wind, that the radius of curvature
of an accelerated ion's orbit is 
\begin{equation}
\rho_c  = 2  r_L \gamma_w^2 = (2 \eta \gamma_w^2 /\cos i) r,
\end{equation}
where $r_L = m_i c^2 \gamma /qB = \eta r$ is an ion's formal Larmor radius.
When $\gamma_w \approx r/R_l \gg 1$, as it is at radii comparable to and larger than
$R_a$, $\rho_c \gg r$.

This large radius of curvature of ions' orbits, applicable wherever the flow is relatistically
expanding (both within the wind and in the MWN) means that both synchrotron and adiabatic losses
of the ions are negligible. From the power radiated by a charged particle with orbital radius
of curvature $\rho_c, \; P= (2q^2 c/\rho^2) \gamma^4 $, $\gamma_w \approx r/R_L $,
$\gamma = \eta q \Phi /m_i c^2 $, one readily finds
\begin{equation}
T_{rad} = \frac{6}{\eta \Omega} \frac{A^4}{Z^5} \left(\frac{m_p c^2}{e\Phi}\right)^3 
          \frac{m_pc^2}{e^2} R_L \sec^2 i \left(\frac{r}{R_L}\right)^6 
           = \frac{3.6 }{\eta_1 \Omega_4^2} \frac{A^4}{Z^5}
              \frac{\sec^2 i}{\mu_{33}^3} r_{10}^6 \; {\rm seconds},
\label{eq:Trad}
\end{equation}
where $\eta_1 = \eta/0.1 $ and $r_{10} = r /10^{10} $ cm; one also obtains this same
radiation time by applying ordinary synchrotron losses in the fluid frame of the wind
where ${\bf E} = {\bf 0}$.  Thus, for $r \geq R_a > 10^{10}$ cm,
radiation losses are negligible ($T_{rad} \gg r/c$). Also, since $\rho \gg r $, 
the high energy ions
are not coupled to the field lines and do not suffer adiabatic expansion losses, so long as they
escape the wind and the MWN while the MWN's expansion is relativistic.

The particles leave the wind and move through the MWN 
with speed $v(t) = c(1 - 1/2\gamma^2 ) \approx c$.  
For the observed UHE cosmic rays, with $E_a < E < 2 \times 10^{20} \; {\rm eV} \ll E_{max}$, 
the particles at radii large compared to $R_s$ were accelerated at radius
$R_a \; (R_a < R_s $ by assumption) and at the time
$T_a = R_a /c + (E_a/E) t_a $, where $t_a = (9/8) (Ze\eta /\mu E_a ) 
\simeq 2 Z \eta_1 I_{45}/\mu_{33} E_{a,18.8} $ days is the time at which the neutron star's
voltage was at the value corresponding to acceleration of
ions at the ankle in the spectrum.
Here $E_{a,18.8} \equiv E_a /6 \times 10^{18}$ eV. For particles in the UHECR
spectrum measured to date, $T_a \approx t_a (E_a /E) $, so long as
$R_a \leq R_s$.
At time $t$, particles of energy $E$ have reached $R_{particle} = c(t-T_a) + R_a$; for
all particles of energy $E >E_a$, $\gamma = E/mc^2 \gg \Gamma_b (t)$.

The outer boundary of the MWN lies at $R_b (t)$ given by (\ref{eq:blast_radius}). All
particles with energy  $E \geq E_a $ eV catch up with the
blast wave at the time $T_{catchup}$ 
when $R_{particle} $ first exceeds $R_b (t)$. Equating the radii yields
\begin{equation}
T_{catchup} =\left(\frac{17}{\pi}\frac{\Delta E_{EM}}{\rho_1 c^5} T_a \right)^{1/4} =
       8.3 \left( \frac{\Delta E_{EM,52}}{n_1} \frac{I_{45}}{\mu_{33} E_{a,18.8}}
            \frac{E_a}{E} \right)^{1/4} 
       {\rm months}.
\label{eq:catch_up_time}
\end{equation}
Here $\Delta E_{EM,52} = \Delta E_{EM} /10^{52}$ ergs.

Comparison of (\ref{eq:catch_up_time}) to (\ref{eq:Trel}) shows that all the UHE
cosmic rays reach the blast wave before the MWN becomes nonrelativistic.
The 4 velocity of the blast wave at $T_c$ is 
$\beta_b \Gamma_b(T_c) = 4 (E_{52}/n_1)^{1/8} (\mu_{33} E_{a,18.8}/Z\eta_1 I_{45})^{3/8}
    (E/E_a)^{3/8}$.
The cosmic rays have no difficulty getting through the plowed up interstellar
medium and its compressed magnetic field just behind the blast wave, once they
arrive at the MWN's outer boundary - the cosmic rays' Larmor radii in the
shocked interstellar magnetic field are
$r_{L2} = 540 (E_{a,18.8}/B_{1,\mu}) (4/\Gamma_b[T_c(E_a)]) (E/E_a)^{5/8}$ parsecs,
which is far larger than the
thickness of the shocked shell of interstellar matter. Thus the UHE ions avoid substantial
adiabatic energy loss in escaping the MWN. Here $B_{1,\mu} = B_{ISM}/1\,\mu{\rm Gauss}$.

Other radiation losses in the MWN might occur if the high energy ions
encounter a sufficiently dense external radiation field.  The supernova's own optical and
ultraviolet luminosity is the most prominent candidate for such radiation.  A reasonable
upper limit for that luminosity is $\sim 10^{44}$ ergs/sec, emitted over the first few days,
{\it i.e.}, over the same time as the UHECR acceleration occurs in the wind.  At the
radii where the acceleration is likely to occur (see \S \ref{sec:accel}), 
photopion losses in the stellar radiation field might limit the maximum ion energy
that can be achieved to no more than $10^{22}$ eV. Losses due to Compton scattering
and pair creation are negligible, as is energy loss in
the dense matter of the supernova, since the ejecta have only a small
filling factor, and move out slowly {\it behind} the high energy particles. 
Losses in the radiation field of the MWN are unlikely to be of significance, since this 
highly magnetized flow is unlikely to be radiatively dissipative until after the
outer blast wave becomes nonrelativistic, which occurs after the UHE ions have left the system.
Radiation losses in the interstellar
environment surrounding the MWN (pion and pair production, synchrotron
radiation, inverse Compton scattering) are all negligible, even if the MWN forms
within a cluster of hundreds of O and B stars, and the magnetic field strength is
that of molecular clouds (milligauss). Interactions with the circumstellar matter that
might surround the presupernova star also can be neglected.

Radiationless escape of the UHE ions from the MWN works only when the ions 
form the equatorial return current. This geometry applies to rotators with dipole
axes making an acute angle with respect to the rotation axis.  Thus, only
50\% of the magnetars can supply UHE cosmic rays, contributing the factor
$W_{geom} = 0.5$ to the flux of UHECR (expression \ref{eq:spectrum}).

If the relativistic wind has a dense pair plasma, as well as the ion return current in the equator
and the polar electron current,
Compton scattering from the pairs might create a high energy photon flash, which may
be an interesting candidate for a prompt Gamma Ray Burst.  This is a topic for a separate
investigation. For the present purposes, it suffices to observe that the supernova photon 
luminosity is small compared to the magnetar's spin down luminosity, therefore Compton drag
can have little effect in dissipating the electromagnetically dominated outflow.

\section{Interstellar Supershells and Supernova Remnants
   \label{sec:supershells}}

Formation of interstellar HI shells and supershells provide an interesting
constraint pointing toward only a small fraction of the newly born magnetars 
having voltages large enough to accelerate UHECR, and perhaps to large gravitational wave emission 
by such objects.

If the magnetars lost energy purely electromagnetically, they 
would deliver energy on the order of
$(1/2) I \Omega_i^2 W_{blowout} = 3.5 \times 10^{52} I_{45} \Omega_4^2 (W_{blowout} /0.7)$ 
ergs with each event.  These explosions would generate supershells, large expanding structures
in the interstellar medium with energy in excess of $10^{52}$ ergs (Heiles 1979, 1984) and
dimensions in excess of 100 pc. From
Chevalier's (1974) expression (26), with a random velocity of clouds in the ISM of $\sim 20$
km/s, electromagnetic energy deposition with $\Omega_4 \sim 1$
would create shells with radii $\sim 230$ pc.
Further investigation over the last 20 years has led to the conclusion that {\it all} such
events in our galaxy occur at a rate 
$\nu_{supershell} \sim 0.8 \times 10^{-5} \; {\rm yr}^{-1}$ (Ehlerova and Palous 1996),
about a factor of 12 slower than the current estimate of the magnetar birth rate; other relations
between explosion energy and remnant size ({\it e.g.} McKee and Cowie 1975) lead to similar
conclusions. Shells as large as 200
pc form only a small, but ill determined, fraction of the total supershell population.
Therefore, if the theory outlined here is to be a viable explanation of UHECR,
either the magnetar rate is overestimated, or the energy per magnetar delivered to the
interstellar medium is smaller (or both).

The approximate correspondence shown 
in Figure 1 between the theory outlined in this paper and the observed
UHECR spectrum  
requires $K_0 = W_{geom} \nu_{m4} (I_{45}/\mu_{33})n_{g2} T_{H14} \approx 0.03 $ if the
spindown is purely electromagnetic.  This result comes from requiring the model
to explain the  well determined spectrum between $10^{19}$ and  $6 \times 10^{19} $ eV. If
the particle spectrum measured at energies below the ankle extrapolates into the
UHE regime, $K_0$ drops to 0.02 for the pure electromagnetic case. If one restricts oneself
to magnetars with parameters defined by current observations, the main freedom is in the 
fact that we don't know whether all such objects are born with the same initial rotation rate.
Magnetars can succeed as a model for UHECR if their voltages are sufficiently large,
requiring initial periods $P_i < 4 (\eta_1 \mu_{33} )^{1/2}$ milliseconds for at least
some of the magnetar population.  

With $W_{geom} = 0.5$,
and $n_g / 0.02 {\rm Mpc}^{-3} \sim 1$ , 
the inferred value of $K_0$ tells us that the birth rate of such rapidly rotating objects, 
which by hypothesis are subject only to electromagnetic torques, is 
$\nu_m^{(fast)} \approx 0.1 \nu_m =
10^{-5} \nu_{m4} \; {\rm yr}^{-1}$ (see \S \ref{sec:predicted}). Then 
most of the
supershells in our galaxy would be the result of these
``electromagnetic bombs'' in the interstellar medium, and most would be of the largest
variety. Loeb and Perna (1998) made the related suggestion that Gamma Ray Bursts drive supershells. 

If one insists that most supershells do not come from exotic sources (probably true), and 
that their numbers not be dominated
by the largest variety (certainly true), then
the energy delivered to the interstellar
medium in each event is less than the $(1/2) I\Omega_i^2 = 5 \times 10^{52} I_{45} \Omega_4^2$ ergs.
For pure electromagnetic spindown,  initial rotation rates less than $10^4$ sec$^{-1}$ are the only
means of achieving that reduction.  Creating shells smaller than 100 pc, which would 
cause them to be missed 
in Heiles's (1984) census, requires the energy deposited per event to be less than 
$4 \times 10^{51}$ ergs, in turn requiring $\Omega_i < 3 \times 10^3 {\rm sec}^{-1}.$
Then the maximum cosmic ray energy would be 
$E_{max} < 2 \times 10^{20} Z\eta_1 \mu_{33} (W_{blowout}/0.7)^{1/2}$
eV, which is too low, for $Z \sim 1$. 

Furthermore, ordinary supernova remnants in galaxies
appear to have sizes {\it typically} less than $\sim 50$ pc (see Matonick and Fesen 1997 
and references therein).
Therefore, from Chevalier's energy-size relation, the energy per event would have to be 
$\sim 10^{50.6}$ ergs - indeed, the standard result, that supernovae create explosions
with energy $\sim 10^{51} $ ergs, comes from precisely these considerations. Requiring the 
rotational energy per magnetar to be $10^{51}$ ergs or less requires initial angular velocities
no greater than $1.4 \times 10^3 I_{45}^{-1/2} \; {\rm sec}^{-1}$, corresponding to
a maximum cosmic ray energy not greater than $7 \times 10^{19} Z\eta_1 /I_{45}$ eV, which is 
insufficient to explain the highest energy cosmic rays (if $Z = 1 - 2$). 

Therefore, the most plausible conclusion 
is that gravitational radiation must be the recipient of most of the initial rotational
energy. The ultra-strong magnetic fields in these stars make this a viable possibility. From
(\ref{eq:EM_lost}), the energy delivered would then be
$E = (1/2) I \Omega_g^2 \ln \Lambda (\Omega \rightarrow 0) W_{blowout,GR} = 
0.75 \times 10^{52} (\mu_{33}^2 /I_{45} \epsilon_2^2 )(\ln \Lambda /2.4) (W_{blowout}/0.7)$
ergs. Requiring the interstellar shells created by these electromagnetic explosions to all
have radii less than 100 pc demands that the neutron stars have equatorial
ellipticities $\epsilon > 2 \times 10^{-2}$; requiring the interstellar remnants of the
explosions to have sizes less than 50 pc demands $\epsilon > 4 \times 10^{-2}$. From
expression (\ref{eq:mag_ellipt}), such ellipticities would be expected for interior
magnetic fields $> (6-8) \times 10^{16} $ Gauss, large but within the realm
of recent theoretical suggestions.
Then more than 90\% of the initial rotational
energy of the fast rotators radiates away as gravitational waves, 
which still leaves more than enough energy for the UHECR. 

In passing, it is interesting to note that Matonick and 
Fesen (1997) found a small number of unusually large (diameters exceeding 100 pc) supernova remnants 
in their sample, with velocities too large to be readily explained as the consequence of several
ordinary but coeval supernova explosions. Such a species of supernova remnant clearly is a candidate
for identification with the electromagnetically driven explosions discussed here.  
However, given the selection effects, it is unclear
whether the number of these unusual events is consistent with electromagnetically driven explosions
from magnetars (or any other exotic explanation) as their origin.  

\section{Gravitational Radiation \label{sec:gravity_waves}}

Since this model requires significant influence on the magnetars' spindown by
gravitational radiation losses, with
release of most of the initial rotational energy going into this so far undetected carrier, the 
model makes reasonably definite predictions for the detection of 
almost coherent signals by LIGO, VIRGO and other gravity wave detectors
optimized for khz signals. The dimensionless strain of the narrow band oscillatory signal,
measured at distance $D$ from the source and
observed for a time longer than the coherence time of the oscillator $\tau_{coh} $, so that
$n \approx \Omega \tau_{coh} $ oscillations can be coherently counted, 
is (Thorne 1997, Brady {\it et al.} 1998)
\begin{equation}
h_n = 2.9 \frac{G  I \epsilon \Omega^2}{D c^4} \sqrt{\Omega \tau_{coh}} ;
\label{eq:strain}
\end{equation}
for simplicity, I have averaged over all orientations of the rotation axis with
respect to the observer, and I have included only the wave modes at the star's rotation 
frequency, which exist
for magnetically distorted stars (Bonazzola and Gourgoulhon 1996).  The coherence time
is $\tau_{coh} = \Omega / | \dot{\Omega} |$.  

Since gravitational radiation
controls the early spindown ($\tau_{EM} \gg \tau_{GW}$),
$\tau_{coh} = 4\tau_{GW} [1 +(t/\tau_{GW})]$ and $\Omega = \Omega_i /(1+ t/\tau_{GW})^{1/4}$
for $t < 2\tau_g = 1.8 I_{45}^3 \epsilon_2^2 /\mu_{33}^4$ hours.
Then 
\begin{eqnarray}
h_n^{(GR)} & \approx & 4 \times 10^{-24} 
   \frac{I_{45} \epsilon_2 \Omega_4^2}{D_{20} [1+(t/\tau_{GW})]^{1/2}} \sqrt{\Omega \tau_{coh}}
= \frac{(\Omega_4 I_{45} )^{1/2}}
      {D_{20} \left( 1 + \frac{t}{\tau_{GW}} \right)^{1/8} } \nonumber \\
    & = & 2.5 \times 10^{-21} \frac{I_{45}^{3/8}}{D_{20} \epsilon_2^{1/4} t_{hours}^{1/8}};
\label{eq:GR_strain}
\end{eqnarray}
here $D_{20} = D/20$ Mpc. For $t > 2 \tau_g $ (rotation periods exceeding
4 msec), 
the electromagnetic torques take over and
\begin{equation}
h_n^{(EM)} = 6.5 \times 10^{-25} \frac{I_{45}^2}{D_{20}} 
      \frac{\epsilon_2}{\mu_{33}^2 t_{hours}} \sqrt{\Omega \tau_{coh}}
  = 2.9 \times 10^{-21} \frac{I_{45}^{9/4} \epsilon_2 }{D_{20} t_{hours}^{3/4}}.
      \label{eq:EM_strain}
\end{equation}

These strain levels are potentially observable, although the problems of finding
such quasi-coherent signals in blind searches, with no advance notice of the period,
are quite formidable.  Nevertheless, since the site of the cosmic ray acceleration
is at small enough radii ($r < R_s \sim $ 1 AU), the UHE cosmic rays from an individual 
outburst should arrive before the gravity waves fade from view - conceivably, a UHE 
cosmic ray event, if observed with sufficient collecting area to identify a burst of
particles, might serve as a marker for searches of gravity wave data for quasi-coherent 
signals with periods in the one to few millisecond regime.

\section{Individual Events \label{sec:beaming}}

Charged particles of energy in excess of $10^{20}$ eV must come from nearby. For
such ultra-high energy particles, the GZK losses
limit the observable volume to size 
\begin{eqnarray}
D_E & = & \min \left\{ D_{GZK} \left[\max \left(\ln \frac{E_{max}}{E},\frac{E_G^2}{E^2} 
       \right) \right], \: D_{H\pm} \right\}, \label{eq:dist_to_E} \\
D_{H\pm} & = & \min \left[ cT_H, \: l_\pm \left(\frac{E_\pm}{E} \right)^{0.4} \right].
\nonumber
\end{eqnarray}
$D_E \geq 50$ Mpc, for maximum energies appropriate to the model discussed here. 
See expression (\ref{eq:loss-rate}) and the immediately preceeding discussion for 
the values of the characteristic lengths and energies.  The number of source events 
per unit time which we can detect, in principle,
in the ultra-high energy particle spectrum then is
$\dot{N}_{UHE} = (4 \pi /3) D_E^3 n_g \nu_m^{(fast)} $. This rate is $\sim$ eight thousand per year
at $E_a = 6 \times 10^{18}$ eV ($D_E \approx 2.1$ Gpc), drops to 800/year at $4 \times 10^{19}$ eV
($D_E \approx 1$ Gpc), but then declines
rapidly, at a rate $\propto E^{-6}$ to a few times per decade above $3 \times 10^{20}$ eV,
assuming the fiducial event rate per unit volume 
$n_g \nu_m^{(fast)} = 2 \times 10^{-7} $
Mpc$^{-3}$ yr$^{-1}$. The particle fluence per event is
\begin{equation}
{\mathcal F}(>E) = \frac{9}{32 \pi D^2} \frac{c^2 I}{  Ze\mu}  f(E)
 \sim 4 \times 10^{-2} \frac{I_{45}}{Z \mu_{33} D_{100}^2} \frac{f}{2.5}\; {\rm km}^{-2};
\label{eq:fluence}
\end{equation}
here $D_{100} = D/100$ Mpc. $f \sim 2-3$ is a logarithmic function of $E$ and $D$.

The model proposed here requires the particles to be accelerated and emitted 
in an equatorial slice with opening angle $2i$, with $i$ the angle between
the rotation axis and the magnetarÕs magnetic moment, into the surrounding Universe - 
the particles are physically beamed.  The fraction of the
sky into which the particles are beamed is $b = \sin i $. If $i$ is uniformly 
distrubuted between $0$ and $\pi /2$, the average
beaming fraction is $\langle b \rangle = 2/\pi = 0.64$.  The individual fluence/event 
would be higher by a factor $1/b$.  Of course, the rate at which events are observed is lower
by a factor $b$, which leads to the average spectrum shown in Figure \ref{fig:theory_data} 
being independent of $b$.

Occasionally (once in $1/\nu_m^{(fast)} \sim 10^5 $ years), such events must occur in our own galaxy,
which would create a primary particle fluence at the earth of order $10^7 $ km$^{-2}$, for 
an average distance of  5 kpc. Such events might leave tracers in the geochemical record, a topic
beyond the scope of this study.

\section{Discussion: Related Models \label{sec:other_models}}

Blasi {\it et al.} (2000) presented the model most closely related to the
theory outlined here.  They were the first to identify the relativistic 
wind of a magnetized
rotator as a potential site of UHE cosmic ray acceleration. 
They assumed the accelerators to be initially rapidly rotating pulsars 
in our own galaxy ($\Omega_i \sim 3000 \; {\rm s}^{-1} $),  and made the same assumption of
the charge lost per second, $I = q\dot{N}_i = \mu \Omega^2 /c = c\Phi$. Since they assumed
normal (if rapidly rotating) pulsars, they produced particles with energy $\sim 100$
EeV only by assuming the high energy ions to be fully stripped iron.  They made the 
same construction of the $E^{-1}$
particle spectrum as in \S \ref{sec:injection} above, for the pure electromagnetic
spindown case. They did not include possible gravitational wave losses.
Since the pulsar spin down time for their assumed
parameters is long compared to the expansion time of the supernova envelope, the
pulsar's rotational energy loss does not substantially affect the dynamics of the
supernova, and
most of the UHE particles form at times late enough to avoid substantial
radiation  and stopping losses in the supernova envelope. They gave no consideration
to the disruption of the envelope due to the pulsars' magnetic pressure on the 
expelled envelopes. They did not use supershells and the sizes of supernova remnants, to
constrain the event rate. The large charge/particle 
assumed allowed them to argue that scattering in the galactic magnetic field might allow 
this hypothesized source to be in reasonable accord with the observed UHECR isotropy. 

However,
UHECR observations support light nuclei, rather than Fe, as the particles responsible for
the UHE showers ({\it e.g.}, Bird {\it et al.} 1994, Watson 2002). 
X-ray observations of pulsars suggest
the atmospheres from which the ions would be extracted are primarily H or He,
as discussed in \S \ref{sec:background}, not Fe. The energies of protons or $\alpha$-particles
would be too low to explain the super-GZK particles in their model. Also,
the Larmor radii would be too large, to allow isotropization in the galactic magnetic field. 
Finally, within the scheme actually outlined
by Blasi {\it et al.}, with each pulsar's magnetic field contained within a closed shell of
supernova ejecta, the supposed UHE cosmic rays suffer catastropic adiabatic losses. 
Because of the relatively small rotational energy per pulsar assumed in their model,
and the slow release of that energy ($\tau_{EM} \sim 1$ year for their parameters),
disruption of the shell and escape of the wind and its high energy particles takes
too long for the UHE particles to survive, with ordinary (if rapidly rotating) pulsars as
the drivers.

The metagalactic magnetar model outlined in this paper avoids all 
of these difficulties, 
as well as adding consideration of the specific mechanism of acceleration in the wind and of
gravitational wave losses on the rotation history
of the pulsar, which can have observable effects on the UHE particle spectrum - see
Figure \ref{fig:theory_data}.

The theory of supernova envelope disruption and relativistic expansion of a Magnetar Wind 
outlined in \S\S \ref{sec:escape} - \ref{sec:blast} have close kinship, and owe much to, the
Gamma Ray Burst models of Wheeler {\it et al.} (2000) and of Inoue {\it et al.} (2002).
The principal difference from the Wheeler {\it et al.} scenario is in the disruption of the 
supernova envelope by Rayleigh-Taylor instabilities, rather than assuming a jet punches 
through the envelope.  Inoue {\it et al.} asummed a rather lower energy rotator, similar
to Blasi {\it et al.'s} fiducial objects, so that Rayleigh-Taylor disruption of the ejecta
happens at rather later times. They also assumed the Pulsar Wind Nebula in their model
would be radiatively efficient so as to construct a Gamma Ray Burst model based on 
nonthermal emission from the Pulsar Wind Nebula, even though they 
also assumed the outflow to be
in the form of an electromagnetically driven relativistic blast wave, an outlow which may be
an inefficient radiator, at least until it decelerates to nonrelativistic velocities. In the present model,
while there may be an initial flash of Compton scattering as the electromagnetic
fields disrupt the supernova shell, the subsequent flow is highly magnetized. Such flows
have weak shock dissipation (Kennel and Coroniti 1984), and may have weak magnetic dissipation
(Lyubarsky and Kirk 2001), so the conversion of outflow energy
to efficiently radiating accelerated particles is likely to be weak.  However, other
processes in a strongly magnetized medium associated with intense electric current flow
may allow conversion of magnetic energy to emergent photons from the wind with unknown efficiency
- see Blandford (2002) for comments on some of the possibilities.

The considerations of laminar surf-riding acceleration outlined in \S \ref{sec:accel} 
have their roots in the work of Buckley (1977) and of Contopoulos and Kazanas (2002).
The relation of these accelerating, force-free wind results to MHD analyses of relativistic
winds will be discussed elsewhere.

Magnetars behaving as outlined in this paper probably cannot co-exist with long lived 
($t_{disk} \gg \tau_{GW}$) fall back
disks, as are invoked in collapsar and supranova models of GRBs - such structures would
probably suppress or greatly modify the flow and acceleration of the ion return
currents in the rotational equator. However, if that disk is itself strongly magnetized,
in principle it could play the same role as a UHECR accelerator.
The open magnetic flux in the magnetar model, $\Psi = R_L \Phi \sim 10^{29} $ Maxwells, is
similar to that expected in many models of Gamma Ray Burst sources' disks, either due to binary
mergers or to fallback in a supernova (Narayan {\it et al.} 1992, Usov 1994,
Paczynski 1998, Vietri and Stella 1999, Macfadyen and Woosley 1999). Many of these models
involve magnetic fields with magnitude $> 10^{16}$ Gauss 
rotating at Keplerian angular velocity at $r \geq 6 R_{Schwarschild} = 179 M/10\,M_\odot$
km. These objects would have $\Psi \sim 10^{30}$ Mx and $\Phi \sim 10^{24.7} $ Volts, clearly
making them available as candidates for UHECR acceleration. Time dependent decay of the disk
might play a role similar to a magnetar's spindown, in creating a broad energy spectrum of high
energy particles.  Such a model has the advantage that
the $>$ petagauss magnetic fields in the disks, if both the disks and the strong fields
exist, are surely rapidly rotating. The magnetar model has the advantage  
that neutron stars with petagauss surface fields do exist, but we do not have independent 
evidence on how many are born rapidly rotating.

\section{Conclusions \label{sec:conclusions}}

The model's basic results and predictions are shown in Figure \ref{fig:theory_data}.  The upper
cutoff of the spectrum, at $E = 10^{21.5} Z\eta_1 \mu_{33} \Omega_4^2 $ eV, is likely
to be reduced by dissipation of the wind as it first breaks free of
the magnetar's natal supernova.  The estimates
of \S \ref{sec:escape} suggest that escape from the supernova consumes a 
fraction $1 - W_{blowout} \sim 0.3 - 0.4$ of the
magnetar's intial energy loss, in turn suggesting that $\sim 80$\% of the highest voltages
are available for cosmic ray acceleration.  If so, then the highest energies in the spectra
shown in Figure \ref{fig:theory_data} should be be taken seriously. 

Agreement with the well determined
UHE cosmic ray flux below $5 \times 10^{19} $ eV constrains the combination
$W_{geom} n_{g2} \nu_{m4}/\mu_{33}$ to be approximately 0.02-0.06, where $\mu_{33}$ is 
the magnetic moment in
units of $10^{33}$ cgs, $\nu_{m4}$ is the magnetar birth rate in units of $10^{-4}$ 
yr$^{-1}$, $n_{g2} = n_{galaxy}/0.02 \; {\rm Mpc}^{-3}$ and $W_{geom} = 0.5$ is 
the inefficiency factor due to the fact that only 50\% of the magnetic geometries
are appropriate to the ion current being emitted into the rotational equator, where
radiation and adiabatic losses on the UHECR would be weak.  The small value of this normalization
required to get a fit, appropriate to the ``no GR'' and ``moderate'' GR cases considered
in Figure 1, implies that the birth rate of magnetars with voltages
high enough to create UHECR is 5-10\% of the overall magnetar birth rate, a conclusion consistent
with not overpopulating our and other galaxies with too many supershells and overly large
supernova remnants. 

The high energy ions 
suffer negligible adiabatic
and radiation losses in escaping the magnetars' winds and their surrounding, relativistically
expanding nebulae.
The existing observations require voltages $\Phi > 2 \times 10^{21} /Z\eta_1 $ Volts,
or $\mu_{33} \Omega_4^2 > 0.06 /\eta_1 Z \; (P_{initial} < 2.6 (Z\eta_1 \mu_{33})^{1/2}$ ms),
with $\eta_1 = \eta /0.1 $ the fraction of the 
voltage actually sampled by each ion. I suggested in \S \ref{sec:accel} 
that electromagnetic surf-riding in the
relativistically strong electromagnetic waves in the wind, generated by the oblique
rotator, is responsible for the ion acceleration, occuring at or within the radius ($\ll 0.1$ AU) 
where the wave structures no longer are frozen into the expanding plasma. That acceleration
site is consistent with small radiative losses.

This model works only if the ions travel from their source galaxies on more or less straight
lines. That requires intergalactic magnetic turbulence to be small amplitude, or the intergalactic
magnetic field to be weak, or both - 
$\delta B /B  < 10^{-2} (10^{-9} \; {\rm Gauss}/ Z B_{IGM})^{1/2}$
(\S \ref{sec:scatter}). Such weak scattering introduces negligible time delays into the
particles' transport from source to observer.

If the current data are interpreted as providing evidence for the conventional GZK
cutoff, {\it i.e.} the AGASA results are disregarded (Bahcall and Waxman 2003) or are
renormalized to bring them into accord with the Hi-Res results (Abu-Zayyad {\it et al.} 2002b),
agreement of this model with existing observations requires rather strong
losses of rotational energy due to gravitational radiation. The ``strong GR''
curve in Figure (\ref{fig:theory_data}) shows the effect of the rapid spindown 
depopulating the highest energy end of the injection spectrum.  The model 
can replicate the high energy end of the spectrum observed by AGASA only if gravitational
wave losses are small. The metagalactic magnetar model has no trouble with a 
spectrum that compromises between the highest energy AGASA results and the Hi-Res (without
the 1995 Bird {\it et al.} event) results - including that data point favors such a compromise.

Even when one assumes a birth rate of fast magnetars $\sim$ 5-10\% of the overall magnetar birth
rate, the requirement that the model not overproduce supershells and large supernova remnants
in our and other galaxies' interstellar
media suggests gravitational wave losses are an important drain of rapidly rotating magnetars' initial
rotational energy. Therefore, one expects bursts of almost coherent gravitational waves
with millisecond periods and strains with magnitude $\geq 10^{-21}$ if observed in an
experiment designed to find almost coherent signals, lasting for several hours 
and overlapping the arrival of the higher
energy particles.

The model suggests the UHE cosmic rays come from sources whose distribution
should mimic that of luminous baryons. More specifically, the sources should follow
luminous matter with stars that are progenitors of core collapse SNe, thus should anticorrelate
with large galaxy clusters. Events with energy above $10^{20}$ eV come from 
small distances ($D < 50$ Mpc), which may allow some imprint of galaxies' large scale structure
to appear on the isotropy of such events. Starburst galaxies are obvious suspects for especially
luminous UHECR sources.

In conclusion, I have shown that bare magnetars in normal galaxies provide a possible source for the
origin of UHECR, both in total flux and in form of the spectrum. I have also given 
plausible arguments for how the particles accelerated by a magnetar embedded in its
supernova and its magnetar wind nebula can escape without catastrophic losses, to contribute
to the particle spectrum observed at the Earth.

This subject will remain observationally driven, most prominently by the 
AGASA ({\it e.g.} Takeda {\it et al.} 1998), Hi-Res ({\it e.g.}
Sokolosky 1998, Abu-Zayyad {\it et al.} 2002b) and Auger ({\it e.g.} Boratov 1997, Cronin 2001a,b) 
experiments, which provide a window into the spectrum at energies above $10^{20.5} $ eV.
The theory described here suggests that there will be something to see at these super-GZK
energies. Also,
Auger, with several thousand km$^2$ collecting area and with fluorescence and particle detectors
recording events simultaneously at the same site, will have the opportunity to
resolve individual UHE particle bursts, with enough statistics to measure a spectrum for
one event, at least up to $10^{20}$ eV - see \S \ref{sec:beaming} for bookeeping
on this model's predictions. On the theoretical front,
the theory of how the acceleration actually occurs, in the model presented here
and in other schemes, urgently needs substantial improvements.

\section{Acknowledgments}

I am indebted to L. Bildsten, E. Brown, G. Farrar, A. Filippenko, C. Heiles, Y. Levin, C.-P. Ma, 
C. Max,  C. McKee, A. Olinto, P.B. Price, S. Singh, 
A. Spitkovsky, and S. Woosley for enlightening discussions. I acknowledge partial support 
from NASA grants
NAG5-12031 and HST-AR-09548.01-A, and from the Miller Institute for Basic Research in Science. 
As always, I thank California's taxpayers for their indulgence.

\appendix
\section{Appendix - R-mode Gravitational Wave Losses\label{sec:app}}

Gravitational wave losses modify the accelerated spectrum by shortening the time each star
spends at a given rotation frequency, thus reducing the flux at energy $E(\Omega)$.  I used
gravitational wave spindown rates owing to permanent nonaxisymmetric quadupolar distortions 
of the stars in order to arrive at the spectra shown in Figure 1, and
argued that these could be generated by strong internal magnetic fields.  The R-mode
instability (Andersson 1998) is another possible source of gravitational wave emission from a 
rapidly rotating neutron star.  With a dimensionless mode amplitude $\alpha$, the spindown
rate due to gravitational radiation is (Owen {\it et al.} 1998)
\begin{equation}
-\dot{\Omega}_{Rmode} = \frac{512}{315} \frac{GI \Omega^5}{c^5} 
      \left(\frac{\Omega R}{c} \right)^2 \alpha^2 ,
\label{eq:Rmode_loss}
\end{equation}
where $R$ is the stellar radius.
The same calculation as in \S \ref{sec:spectrum}, with R mode losses replacing those due to a static
quadrupole moment, leads to
\begin{equation}
\frac{d N_i}{d\gamma} = 
  \frac{d N_i}{dt}\left(-\frac{dt}{d\Omega}\right)mc^2 \frac{d\Omega}{dE} =
  \frac{9}{4} \frac{c^2 I}{Z e \mu \gamma}
       \left[ 1 + \left(\frac{\gamma}{\gamma_g}\right)^2 \right]^{-1},
\label{eq:Rmode-diff-inj-rate}
\end{equation}
replacing expression (\ref{eq:diff-inj-rate}), {\it assuming} $\alpha =$ constant during the spindown,
as in the model of Owen {\it et al.}. In (\ref{eq:Rmode-diff-inj-rate}), the characteristic
energy above which gravitational wave losses affect the spectrum now is 
$$
E_g = m_ic^2 \gamma_g = \frac{Ze\eta \mu}{c^2} \left(\frac{104}{384} 
            \frac{\mu^2 c^4}{GI^2 R^2 \alpha^2 }\right)^{1/2} 
              = 6 \times 10^{20} \frac{Z\eta_1 \mu_{33}^2}{I_{45} R_{10} \alpha } \; {\rm eV},
$$
with $R_{10} = R/10$ km. If $\alpha$ were large, $E_g (\alpha)$ might be low enough to yield a
steepened high energy injection spectrum and a GZK pileup.

However, recently Arras {\it et al.} (2003) provided the theory for $\alpha$ as a function of $\Omega$, 
showing that damping
in the neutron star's interior, through coupling between R modes and internal g modes, keeps
$\alpha $ small. In the notation used here, their theory provides $\alpha = 0.025 \Omega_4^{5/2}$.
The gravitational wave spindown rate is now proportional to $\Omega^{12}$, instead of 
$\Omega^5$ (static quadupole) or $\Omega^7$ (Rossby modes with constant $\alpha$.)
The injection spectrum still is given by (\ref{eq:Rmode-diff-inj-rate}), but now with 
$(\gamma /\gamma_g)^2$ replaced by $(\gamma /\gamma_g)^{9/2}$, and with
$ E_g = m_ic^2 \gamma_g = 10^{21} Z\eta_1 \mu_{33}^{13/9} /(I_{45} R_{10})^{4/9}$ eV.  
Gravitational wave losses due to the R mode instability therefore have negligible effect
on the ability of the magnetar model to account for {\it existing} UHECR observations. They
do suggest a very hard cutoff of the spectrum, above $10^{21}$ eV.


\begin{thebibliography}{}
\bibitem[Abu-Zayyad et al.(2002a)]{abu02a} Abu-Zayyad, T., Archbold, G., Bellido, J.A., {\it et al.}, 2002a, 
  submitted to \prl  (astrop-ph/0208243)
\bibitem[Abu-Zayyad et al.(2002b)]{abu02b} Abu-Zayyad, T., Archbold, G., Bellido, J.A., {\it et al.}, 2002b, 
  submitted to Astroparticle Phys.  (astro-ph/0208243)
\bibitem[Adams et al.(1997)]{ada97} Adams, F.C., Freese, K., Laughlin, G., {\it et al.} 1997, \apj, 491, 6
\bibitem[Andersson(1998)]{and98} Andersson, N. 1998, \apj, 502, 708
\bibitem[Arons and Lea(1981)]{aro81} Arons, J., and Lea, S.M. 1980, \apj, 235, 1016
\bibitem[Arons(2002)]{aro02} Arons, J. 2002, in `Neutron Stars and Supernova RemnantsÕ,  
      P.O. Slane and B.M. Gaensler, eds. (San Francisco: Astronomical 
      Society of the Pacific), 71 (astro-ph/0201439)
\bibitem[Arras(2003)]{arr03} Arras, P., Flanagan, E.A., Morsink, S., {\it et al.}, 2003, \apj, in press
   (astro-ph/0202345)
\bibitem[Baring(2001)]{bar01} Baring, M., and Harding, A.K. 2001, \apj, 547, 929
\bibitem[Bahcall and Waxman(2003)]{bah03} Bahcall, J.N., and Waxman, E. 2003, Phys. Lett. B, 
   556, 1 (hep-ph/020621)
\bibitem[Barrow(1997)]{bar97} Barrow, J.D., Ferreira, P.G., and Silk, J.I. 1997, \prl, 78, 3610 \bibitem[Berezinsky et al.(2002)]{ber02} Berezinsky, V., Gazizov, A.Z. and Grigorieva, S.I. 2002, 
   (hep-ph/0204357)
\bibitem[Bellan(1999)]{bel99} Bellan, P. 1999, \prl, 83, 23
\bibitem[Bellan(2001)]{bel01} Bellan, P. 2001, Adv. Space. Res., 28, 729
\bibitem[Beuermann et al(1985)]{beu85} Beuermann, K., Kanbach, G., and Berkhuijsen, E.M. 1985, \aap, 153, 17
\bibitem[Bird et al. 1994]{bir94} Bird, D.J., Corbato, S.C., Dai, H.Y., {\it et al.} 1994, \apj, 424, 491
\bibitem[Bird et al. 1995]{bir95} Bird, D.J., Corbato, S.C., Dai, H.Y., {\it et al.} 1995, \apj, 441, 144
\bibitem[Blandford and McKee(1976)]{bla76} Blandford, R.D., and McKee, C.F. 1976, Phys. Fluids, 19, 1130
\bibitem[Blandford(2002)]{bla02} Blandford, R.D. 2002, in "Lighthouses of the Universe", 
    ed. M. Gilfanov {\it et al.}  (Berlin:Springer), in press (astro-ph/0202265)
\bibitem[Blanton et al(2001)]{bla01} Blanton, M.R., Dalcanton, J., Eisenstein, D., 
     {\it et al.} 2001, \aj, 121, 2358
\bibitem[Blasi et al.(2000)]{bla00} Blasi, P., Epstein, R.I., and Olinto, A.V. 2000, \apjl, 533, L123
\bibitem[Bogovalov(1997)]{bog97} Bogovalov, S.V. 1997, \aap, 327, 662
\bibitem[Bogovalov(1999)]{bog99} Bogovalov, S.V. 1999, \aap, 349, 1017
\bibitem[Bonazzola and Gourgoulhon(1996)]{bonn96} Bonazzola, S., and Gourgoulhon, E. 1996, \aap, 312, 675
\bibitem[Boratov(1997)]{bor97} Boratov, M. 1997, in Proceedings of the 25th International Cosmic Ray Conference, 
     M. S. Potgieter, B. C., Raubenheimer, and D. J. van der Walt, eds. (Singapore:
     World Scientific), 5, 205
\bibitem[Brady et al. 1998]{bra98} Brady, P.R., Creighton, T. Cutler, C., and Shutz, B.F. 1998, 
     \prd, 57, 2101
\bibitem[Buckley(1977)]{buc77} Buckley, R. 1977, Nature, 266, 37
\bibitem[Cappellaro(1999)]{cap99} Cappellaro, E., Evans, R., and Turatto, M. 1999, \aap, 351, 459
\bibitem[Chen(2002)]{che02} Chen, P., Tajima, T., and Takahasi, Y. 2002, \prl, 89, 161101 (astro-ph/0205287)
\bibitem[Chevalier(1974)]{che94} Chevalier, R. 1974, \apj, 188, 501
\bibitem[Contopoulos et al(1999)]{con99} Contopulos, I., Kazanas, D., and Fendt, C. 1999, \apj, 511, 351
\bibitem[Contopoulos and Kazanas(2002)]{con02} Contopoulos, I., and Kazanas, D. 2002, \apj, 566, 336
\bibitem[Coroniti(1990)]{cor90} Coroniti, F.V. 1990, \apj, 349, 538
\bibitem[Cronin(2001a)]{cro01a} Cronin, J.W. 2001a, In `Observing Ultrahigh Energy Cosmic 
     Rays from Space and Earth', ed. H. Salazar {\it et al.} (New York: AIP Conf. Proc. no. 566), 1 
\bibitem[Cronin(2001b)]{cro01b} Cronin, J.W. 2001b, Nucl. Phys. B (Proc. Suppl.), 97, 3
\bibitem[Davies and Taylor(1950)]{dav50} Davies, R.M., and Taylor, G.I. 1950, Proc. Roy. Soc. 
      London, 200, 375
\bibitem[Duncan and Thompson 1992]{dunc2} Duncan, M. and Thompson, C. 1992, \apjl, 392, L9
\bibitem[Ehlerova and Palous(1996)]{ehl96} Ehlerova, S., and Palous, J. 1996, \aap, 313, 478
\bibitem[Eilek and Owen(2002)]{eil02} Eilek, J.A., and Owen, F.N. 2002, ApJ, 567, 202
\bibitem[Eltgroth(1972)]{elt72} Eltgroth, P.G. 1972, Phys Fluids, 15, 2140
\bibitem[Farrar and Piran(2000)]{far00}  Farrar, G.R., and Piran, T. 2000, \prl, 84, 3527
\bibitem[Filippenko(2001)]{fil01} Filippenko, A.V. 2001, in ``Young Supernova Remnants'', 
     S.S. Holt and U. Hwang, eds., AIP Conf. Proc. No. 565 (NY: American Institute of Physics), 40
\bibitem[Fodor and Katz(2000)]{fod00} Fodor, Z., and Katz, S.D. 2000, \prd, 63, 023002
\bibitem[Gallant and Arons(1994)]{gal94} Gallant, Y.A., and Arons, J. 1994, \apj, 435, 230
\bibitem[Gaensler et al.(2001)]{gaens01} Gaensler, B.M., Slane, P.O., Gotthelf, E.V., and Vasisht, G. 
   2001, \apj, 559, 963
\bibitem[Gaensler et al(2002)]{gae02} Gaensler, B.M., Arons, J., Kaspi, V.M., {\it et al.} 2002, \apj, 569, 
     878 (astro-ph/0110454)
\bibitem[Ginzburg and Syrovatskii(1964)]{gin64} Ginzburg, V.L., and Syrovatskii, S.I. 1964, The Origin of 
     Cosmic Rays (New York:Pergamon), 236-282
\bibitem[Goldreich and Julian(1969)]{gol69} Goldreich, P., and Julian, W.H. 1969, \apj, 157, 869
\bibitem[Greisen(1966)]{gre66} Greisen, K. 1966, \prl, 16, 748
\bibitem[Hailey and Mori 2002]{hai02} Hailey, C.J., and Mori, K. 2002, \apjl, 578, L133  (astro-ph/0207590) 
\bibitem[Halzen and Hooper (2002)]{hal02} Halzen, F., and Hooper, D. 2002, Repts. Prog. Phys., 
    65, 1025 (astro-ph/0204527)
\bibitem[Hayashida et al(1994)]{hay94} Hayashida, N., Honda, K., Honda, M. {\it et al.} 1994, \prl,
    73, 3491
\bibitem[Heiles(1979)]{hei79} Heiles, C. 1979, \apj, 229, 533
\bibitem[Heiles(1984)]{hei84} Heiles, C. 1984, \apjs, 55, 585
\bibitem[Helfand, Gotthelf and Halpern(2001)]{helf01} Helfand, D.J., Gotthelf, E.V., and Halpern, 
   J.P. 2001, \apj, 556, 3801 (astro-ph/0007310)
\bibitem[Hester, Scowen, Sankrit et al.(1995)]{hes95} Hester, J.J., Scowen, P.A., 
    Sankrit, R., {\it et al.} 1995, \apj, 448, 240
\bibitem[Hester, Stone, J.M., Scowen et al(1996)]{hes96} Hester, J.J., Stone, J.M., 
       Scowen, P.A., {\it et al.}. 1996, \apj, 456, 225
\bibitem[Hummel et al(1991)]{hum91} Hummel, E., Dahlem, M., van de Hulst, J.M., and Sukumar, S. 1991,
     \aap, 246, 10
\bibitem[Inoue, Guetta and Pacini(2002)]{ino02} Inoue, S., Guetta, D., and Pacini, F. 2002, \apj, 
    in press (astro-ph/0111591)
\bibitem[Jun, Norman and Stone(1995)]{jun95} Jun, B.-I., Norman, M.L. and Stone, J.M. 1995, \apj, 453, 332
\bibitem[Kaspi and Helfand(2002)]{kasp02} Kaspi, V. and Helfand, D.J. 2002, in `Neutron Stars 
and Supernova RemnantsÕ, ed. P.O. Slane and B.M. Gaensler, (San Francisco: Astronomical 
      Society of the Pacific), 3 (astro-ph/0201183)
\bibitem[Kennel and Coroniti(1984)]{Ken84} Kennel, C.F., and Coroniti, F.V. 1984, \apj, 283, 710
\bibitem[Kouveliotou et al. 1998]{kouv98} Kouveliotou, C. {\it et al.} 1998, \nat, 393, 234
\bibitem[Kouveliotou et al. 1999]{kouv99} Kouveliotou, C. {\it et al.} 1999, \apjl, 510, L115
\bibitem[Kronberg, Dufton, Li and Colgate, S.A.(2001)]{kron01} Kronberg, P.P., Dufton, Q.W., Li, H., 
    and Colgate, S.A. 2001, \apj, 560, 178
\bibitem[Kruskal and Schwarszschild(1954)]{kru54} Kruskal, M., and Schwarzschild, M. 1954, Proc. Roy.
    Soc. London, A223, 348
\bibitem[Kulsrud and Pearce(1969)]{kul69} Kulsrud, R., and Pearce, W.P. 1969, \apj, 156, 445
\bibitem[Lai(2001)]{lai01} Lai, D. 2001, in `Astrophysical Sources for Ground Based Garvitational Wave 
  Detectors', ed. J. Centrella (New York: AIP), 246 (astro-ph/0101042)
\bibitem[Layzer(1955)]{layz55} Layzer, D. 1955, \apj, 122, 1
\bibitem[Leboeuf et al(1982)]{leb82} Leboeuf, J.N., Ashour-Abdalla, M., Tajima, T., {\it et al.} 1982,
   \pra, 25, 1023
\bibitem[Lisenfeld and Volk(2000)]{lis00} Lisenfeld, U., and Volk, H.J. 2000, \aap, 354, 423 
\bibitem[Loeb and Perna(1998)]{loe98} Loeb, A., and Perna, R. 1998, \apjl, 503, L35
\bibitem[Longair(1994)]{lon94} Longair, M. 1994, `High Energy AstrophysicsÕ, vol. 2 (Cambridge:
    Cambridge University Press), 333-342
\bibitem[Lynden-Bell(1969)]{lynd69} Lynden-Bell, D. 1969, \nat, 223, 690 
\bibitem[Lyne and Manchester(1988)]{lyn88} Lyne, A.G., and Manchester, R.N. 1988, \mnras, 234, 477
\bibitem[Lyubarsky and Kirk(2001)]{lyu01} Lyubarsky, Y., and Kirk, J. 2001, \apj, 547, 437
\bibitem[Lyutikov(2002)]{lyu02} Lyutikov, M. 2002, \mnras, submitted  (astro-ph/0210353)
\bibitem[MacFadyen and Woosley(1999)]{macf99} MacFadyen, A.I. and Woosley, S.E. 1999, \apj, 524, 262
\bibitem[Matonick and Fesen(1997)]{mato97} Matonick, D.M., and Fesen, R.A. 1997, \apjs, 112, 49
\bibitem[Mazzali et al(2002)]{maz02} Mazzali, P.A., Deng, J., Maeda, K., {\it et al.} 2002, \apjl, 
     572, L61 (astro-ph/0204007)
\bibitem[McKee and Cowie(1975)]{mck75} McKee, C.F., and Cowie, L.L. 1975, \apj, 195, 715
\bibitem[Melatos(1998)]{mela98} Melatos, A. 1998, Mem. Soc. Ast. It., 69, 1009
\bibitem[Melatos and Melrose(1996)]{mela96} Melatos, A., and Melrose, D. 1996, \mnras, 279, 1168
\bibitem[Michel(1971)]{mich71} Michel, F.C. 1971, Comments on Ap and Space Phys, 3, 80 
\bibitem[Michel(1973)]{mich73} Michel, F.C. 1973, \apjl, 180, L133
\bibitem[Michel(1974)]{mich74} Michel, F.C. 1974, \apj, 187, 585
\bibitem[Michel(1975)]{mich75} Michel, F.C. 1975, \apj, 197, 193
\bibitem[Michel(1994)]{mich94} Michel, F.C. 1994, \apj, 431, 397  
\bibitem[Murray et al.(2002)]{murr02} Murray, S.S., Slane, P.O., Seward, F.D., 
   {\it et al.} 2002, \apj, 568, 226
\bibitem[Nagano and Watson (2000)]{nag00} Nagano, M., and Watson, A.A. 2000, Rev. Mod. Phys., 72, 689
\bibitem[Ostriker and Gunn(1969)]{ostr69} Ostriker, J.P., and Gunn, J.E. 1969, \apj, 157, 1395
\bibitem[Ostrowski and Bednarz(2002)]{ost02} Ostrowski, M., and Bednarz, J. 2002, \aap, in press (astro-ph/0101069)
\bibitem[Owen et al.(1998)]{owen98}Owen, B.J., Lindblom, L., Cutler, C., 
   {\it et al.} 1998, \prd, 58, 084020
\bibitem[Paczynski 1992]{pacz92} Paczynski, B. 1992, Acta Astron., 42, 145
\bibitem[Paczynski(1998)]{pacz98} Paczynski, B. 1998, \apjl, 494, L45
\bibitem[Pavlov and Zavlin (2000)]{pavl00} Pavlov, G.G., and Zavlin, V.E. 2000, in `Pulsar Astronomy - 2000 and Beyond', 
    M. Kramer, N. Wex and R. Wielebinski, eds. (San Francisco: Astronomical Society of the
    Pacific), 613
\bibitem[Protheroe and Johnson(1996)]{proth96} Protheroe, R.J., and Johnson, P. 1996, 
     Astropart. Phys., 4, 253
\bibitem[Ruderman and Sutherland(1975)]{rud75} Ruderman, M.A., and Sutherland, P.G. 1975, 196, 51
\bibitem[Sarkar(2002)]{sark02} Sarkar, S. 2002, preprint (hep-ph/0202013)
\bibitem[Sanwal, Pavlov, Zavlin, and Teter(2002)]{sanw02} Sanwal, D., Pavlov, G.G., Zavlin, V.E., and 
 Teter, M.A. 2002, \apjl, 574, L61  (astro-ph/0206195)
\bibitem[Singh and Ma(2003)]{sing03} Singh, S., and Ma, C.-P. 2003, \prd, 67, 023506 (astro-ph/0208419)
\bibitem[Sokolsky(1998)]{soko98} Sokolsky, P., 1998, in Proceedings of Workshop on Observing 
Giant Cosmic Ray AirShowers from $ E > 10^{20} $ eV Particles from Space, AIP Conf. Proc. No. 433, 
J. F. Krizmanic, J.F. Ormes, and R. E. Streitmatter, eds. (AIP, Woodbury, NY), 65
\bibitem[Spitkovsky and Arons(2000)]{spit00} Spitkovsky, A., and Arons, J. 2000, in `Pulsar Astronomy - 2000 and
    Beyond', IAU Colloquium No. 177 (San Francisco: Astronomical Society of the Pacific), 507
\bibitem[Spitkovsky and Arons(2003)]{spit03} Spitkovsky, A., and Arons, J. 2003, submitted to \apj
\bibitem[Stanev et al.(2000)]{sta00} Stanev, T., Engel, R., Mucke, A., Protheroe, R., and Rachen, J. 
   2000, \prd, 62,  3005
\bibitem[Tajima and Dawson(1979)]{taj79} Tajima, T., and Dawson, J.M. 1979, \prl, 43, 26
\bibitem[Takeda et al(1998)]{tak98} Takeda, M., Hayashida, N., Honda, K., {\it et al.} 1998, \prl,
     81, 1163
\bibitem[Takeda et al. 1999]{tak99} Takeda, M., Hayashida, N., Honda, K., {\it et al.} 1999, \apj, 
     522, 225 (astro-ph/9902239) 
\bibitem[Thorne(1997)]{tho97} Thorne, K.S. 1997, Rev. Mod. Astron., 10, 1
\bibitem[Torres et al.(2002)]{tor02} Torres, D.F., Boldt, E., Hamilton, H., and Loewenstein, M. 2002, 
  \prd, 66, 0203001
\bibitem[Uchihori et al(2000)]{uch00} Uchihori, M., Nagano, M., Takeda, M. {\it et al.} 2000,  
    Astropart. Phys., 13, 151
\bibitem[Ushomirsky(2001)]{ush01} Ushomirsky, G. 2001, in `Astrophysical Sources for Ground Based 
 Gravitational Wave Detectors', ed. J.Centrella (New York: AIP), 284  
\bibitem[Usov(1992)]{uso92} Usov, V.V. 1992, \nat, 357, 472
\bibitem[Usov(1994)]{uso94} Usov, V.V. 1994, \mnras, 267, 1035
\bibitem[Vietri and Stella(1999)]{vie99} Vietri, M. and Stella, L. 1999, \apjl, 527, L43 
\bibitem[Watson (2002)]{wat02} Watson, A.A. 2002, Contemporary Physics, 43, 181
\bibitem[Waxman(1995)]{wax95} Waxman, E. 1995, \prl, 75, 386
\bibitem[Wheeler et al(2000)]{whee00} Wheeler, J.C., Yi, I., Hoflich, P., and Wang, L. 2000, \apj, 537, 810
\bibitem[Yoshida et al. 1995]{yos95} Yoshida, S., Hayashida, N., Honda, K., {\it et al.} 1995, Astropart.
  Phys., 3, 105\bibitem[Zatsepin and Kuzmin(1966)]{zat66} Zatsepin, G.T., and Kuzmin, V.A. 1966, Sov. Phys. - JETP (Lett.), 
    4, 78 
\bibitem[Zavlin, Pavlov and Trumper(1998)]{zavl98} Zavlin, V.E., Pavlov, G.G., and Trumper, J. 1998, \aap, 
   331, 821
\end{thebibliography}
\end{document}